\pdfoutput=1
\documentclass[12pt]{article}
\usepackage{geometry}
\usepackage{amsfonts, amssymb, amsmath}
\usepackage{subcaption}
\usepackage{graphicx, amsthm, multirow}
\usepackage[citecolor=blue]{hyperref}
\usepackage[all]{hypcap}
\usepackage{url}
\usepackage{soul}
\usepackage{xstring}
\usepackage[usenames,dvipsnames]{color}
\usepackage{xspace} 
\usepackage{fancyhdr} 
\usepackage{cite}
\usepackage[font=footnotesize]{caption}

\def\be{\begin{equation}}
\def\ee{\end{equation}}
\def\beq{\begin{equation}}
\def\eeq{\end{equation}}
\def\bea{\begin{eqnarray}}
\def\eea{\end{eqnarray}}

\newcommand{\refref}[1]{Ref.~{\protect\cite{#1}}}
\newcommand{\refsref}[2]{Refs.~{\protect\cite{#1}}~and~{\protect\cite{#2}}}
\newcommand{\figref}[1]{Fig.~{\protect\ref{#1}}}
\newcommand{\secref}[1]{Section~{\protect\ref{#1}}}
\newcommand{\subsecref}[1]{subsection~{\protect\ref{#1}}}
\newcommand{\appref}[1]{Appendix~{\protect\ref{#1}}}

\newcommand{\eref}[1]{Eq.~({\protect\ref{#1}})}

\newcounter{CommentCount}
\definecolor{PB}{rgb}{0,0.5,0}
\definecolor{CL}{rgb}{0,0,1}
\definecolor{MS}{rgb}{1,0,0}
\definecolor{SFK}{rgb}{1,0,1}
\definecolor{SP}{rgb}{0.5,0,0.5}

\newcommand{\gal}[1]{\ensuremath #1_g}

\newcommand{\GRnew}{GR3\xspace} %The GR3 pattern
\newcommand{\TBM}{TBM\xspace} 	%The tribimaximal pattern
\newcommand{\BM}{BM\xspace} 	%The bimaximal pattern
\newcommand{\GR}{GR1\xspace} 	%The original GR
\newcommand{\DGR}{GR2\xspace}	%The dihedral GR
\newcommand{\HEX}{HEX\xspace}	%The hexagonal pattern
\newcommand{\eg}{\emph{e.g.}\xspace}	
\newcommand{\ie}{\emph{i.e.}\xspace}	

\begin{document}
\pagestyle{fancyplain}
\rhead{\fancyplain{IPPP-14-90, DCPT-14-180, SI-HEP-2014-27, QFET-2014-20}{}}
\cfoot{}

%For feynmf-Package:
\setlength{\unitlength}{1mm}

\begin{titlepage}

\title{\bf\Large
Testing solar lepton mixing sum rules in neutrino oscillation experiments}

\author{
Peter~Ballett$^a$\thanks{email: \tt peter.ballett@durham.ac.uk}~,~
Stephen F.~King$^b$\thanks{email: \tt king@soton.ac.uk}~,~
Christoph~Luhn$^c$\thanks{email: \tt christoph.luhn@uni-siegen.de}~,~\\
Silvia~Pascoli$^a$\thanks{email: \tt silvia.pascoli@durham.ac.uk}~,~ and~
Michael A.~Schmidt$^d$\thanks{email: \tt m.schmidt@physics.usyd.edu.au}
\\[6mm]
{\normalsize $^a$ \it IPPP, Department of Physics, Durham University,}\\[-0.5mm]
{\normalsize \it South Road, Durham DH1 3LE, United Kingdom}\\[1ex]
{\normalsize $^b$ \it Physics and Astronomy, University of Southampton,}\\[-0.5mm]
{\normalsize \it Southampton, SO17 1BJ, United Kingdom}\\[1ex]
{\normalsize $^c$ \it Theoretische Physik 1, Naturwissenschaftlich-Technische Fakult\"at,}\\[-0.5mm]
{\normalsize \it Universit\"at Siegen, Walter-Flex-Stra{\ss}e 3, 57068 Siegen, Germany}\\[1ex]
{\normalsize $^d$ \it ARC Centre of Excellence for Particle Physics at the Terascale,}\\[-0.5mm]
{\normalsize \it School of Physics, The University of Sydney, NSW 2006, Australia}\\[4mm]
}

%\date{\today}

\maketitle

~\\[-16mm]\begin{abstract} 
Small discrete family symmetries such as S$_4$, A$_4$ or A$_5$ may lead to
simple leading-order predictions for the neutrino mixing matrix such as the
bimaximal, tribimaximal or golden ratio mixing patterns, which may be brought
into agreement with experimental data with the help of corrections from the 
charged-lepton sector.  Such scenarios generally lead to relations among the
parameters of the physical leptonic mixing matrix known as solar lepton mixing
sum rules. In this article, we present a simple derivation of such solar sum
rules, valid for arbitrary neutrino and charged lepton mixing angles and
phases, assuming only $\theta_{13}^{\nu}=\theta^e_{13}=0$. We discuss four
leading-order neutrino mixing matrices with $\theta_{13}^{\nu}=0$ which are
well motivated from family symmetry considerations. We then perform a
phenomenological analysis of the scope to test the resulting four solar sum
rules, highlighting the complementarity between next-generation neutrino
oscillation experiments such as the reactor experiment JUNO and a superbeam
experiment. 
\end{abstract} 

\end{titlepage}
\setcounter{page}{1}
\fancyhf{}
\pagestyle{plain}
\cfoot{\thepage}

%%%%%%%%%%%%%%%%%%%%%%%%%%%%%%%%%%%%%%%%%%%%%%%%%%%%%%%%%%%%%%%%%%%%%%
\section{\label{sec:Introduction}Introduction}
%%%%%%%%%%%%%%%%%%%%%%%%%%%%%%%%%%%%%%%%%%%%%%%%%%%%%%%%%%%%%%%%%%%%%%

Following the measurement of the third lepton mixing angle, the so-called
reactor angle $\theta_{13}\approx 8.5^{\circ}$ \cite{An:2012eh}, neutrino physics
has entered the precision era. Indeed all three lepton mixing angles are
expected to be measured with increasing precision over the coming years, with
forthcoming accurate measurements expected for both the atmospheric angle
$\theta_{23}$ and the solar angle $\theta_{12}$. First hints of the
CP-violating (CPV) phase $\delta$ have also been reported in global
fits \cite{Gonzalez-Garcia:2014bfa,Forero:2014bxa,Capozzi:2013csa}, and 
rapid progress can be expected with the next generation of oscillation
experiments.  

The measurement of the reactor angle has had a major impact on models of
neutrino mass and mixing, ruling out at a stroke models based on tribimaximal
(TBM) lepton mixing \cite{Harrison:2002er}, although, as we shall discuss in
this paper, these patterns may survive in the neutrino sector, if charged
lepton corrections are included. Such TBM patterns can arise from ``direct'' models
\cite{King:2009ap}, in which the full Klein symmetry ($S,U$ generators) of the
neutrino mass matrix as well as the $T$ symmetry of the charged lepton mass
matrix are subgroups of an underlying discrete family symmetry. Alternatively,
TBM mixing can arise from ``indirect'' models based on constrained sequential
dominance (CSD)~\cite{King:2005bj} with special family symmetry breaking vacuum
alignments. 

In response to the experimental data, many different model building directions
capable of accounting for the reactor angle have emerged, as recently reviewed
in \refsref{King:2013eh}{King:2014nza}.
The viability of these ideas can only be established by comparison with
experiment, and a tractable approach to test large classes of models is to
identify generic types of prediction associated with these models. A promising
example of such a signature can be found in lepton mixing sum rules,
which relate the three lepton mixing angles to the CPV
oscillation phase $\delta$, or more precisely to $\cos \delta$. 
Indeed, given the precisely measured values of the mixing angles,
they can be regarded as predictions for $\cos \delta$, to be tested in future
experiments. Lepton mixing sum rules arise from two distinct types of
scenarios and lead to two different types, referred to as atmospheric and
solar sum rules \cite{King:2013eh,King:2014nza}.

{\em Atmospheric sum rules} \cite{King:2007pr} arise from a variety of
``semi-direct'' models in which only half of the Klein symmetry emerges from
the discrete family symmetry, classified in terms of finite von Dyck groups,
with charged lepton mixing controlled by the $T$ generator
\cite{Hernandez:2012ra,Hernandez:2012sk,Ballett:2013wya}. For example, such
models can lead to trimaximal-1 (TM1) or trimaximal-2 (TM2) mixing, in which
the first or second column of the \TBM mixing matrix is preserved,
{\protect\eref{TM1}} and {\protect\eref{TM2}}
respectively,
\begin{align} &  {\rm TM1}: &\ \  \left | U_{e1} \right | &  =
\sqrt{\frac{2}{3}}\qquad \text{and}\qquad \left|U_{\mu1}\right|
=\left|U_{\tau1}\right| =\frac{1}{\sqrt{6}}\ ;  \label{TM1}\\
& {\rm TM2}: & \ \ \left|U_{e2}\right| & =  \left|U_{\mu2}\right| = \left|U_{\tau2}\right| =
\frac{1}{\sqrt{3}}\ .  \label{TM2} \end{align}
The atmospheric sum rule $a= \lambda r\cos\delta + \mathcal{O}(a^2,r^2)$ can be
derived from these conditions~\cite{King:2007pr}, where
$a\equiv\sqrt{2}\sin\theta_{23}-1$, $r\equiv\sqrt{2}\sin\theta_{13}$ and
$\lambda=1$ for TM1 and $\lambda=-1/2$ for TM2. The study of correlations of
this type, and their application to the discrimination between underlying
models, has been shown to be a realistic aim for a next-generation superbeam
experiment \cite{Ballett:2013wya}.

It was first shown in \refsref{Shimizu:2011xg}{King:2011zj} that A$_4$
generally leads to a ``semi-direct model'' which predicts TM2 mixing with the
second atmospheric sum rule, while the indirect CSD2 model with special
family symmetry breaking vacuum alignments $(0,1,1)^T$ and $(1,2,0)^T$ in
\refref{Antusch:2011ic} predicts TM1 mixing and the first atmospheric sum rule.
In fact the TM1 atmospheric sum rule arises from all generalised versions of
CSD$(n)$, based on the vacuum alignments $(0,1,1)^T$ and $(1,n,n-2)^T$ for integer
$n\geq 1$ \cite{King:2013iva}, since such alignments are orthogonal to the
first column of the TBM matrix, $(2,-1,1)^T/\sqrt{6}$, and hence predict TM1 mixing.

The Pontecorvo-Maki-Nakagawa-Sakata (PMNS) matrix $U$ can be expressed as the
product of the diagonalising matrices of the neutrino and charged lepton mass
terms, $U_\nu$ and $U_e$, respectively, 
\[  U = U^\dagger_e U_\nu.  \]
{\em Solar sum rules} \cite{King:2005bj,Masina:2005hf,Antusch:2005kw} arise in models
in which a leading-order mixing matrix $U_\nu$ is corrected by a small basis
change from the charged leptons.
In the first models of this type $U_e$ had a Cabibbo-like form: if we denote
the angles which parameterise $U_\alpha$ by $\theta^\alpha_{ij}$, the angles
obey $0 \approx \theta^e_{13} \approx \theta^e_{23}\ll\theta^e_{12} \approx
\theta_C$.  These scenarios are motivated by Grand Unified Theories (GUTs)
where the approximately diagonal charged lepton mass matrix is related to the
down-type quark mass matrix, together with the assumption that quark mixing
arises predominantly from the down-type quark sector. Indeed this was the case
in the CSD model where solar sum rules were first proposed \cite{King:2005bj}.
In the context of ``direct'' models, solar sum rules arise when the full Klein
group continues to emerge from the discrete family symmetry (leading for
example to TBM mixing in the neutrino sector) while the $T$ generator which
governs the charged leptons is broken. In the simple case where only
$\theta^e_{12},\theta^{\nu}_{12},\theta^{\nu}_{23}$ are non-zero, with
$\theta^e_{23}=\theta^e_{13}=\theta^{\nu}_{13}=0$, the charged-lepton
corrections do not change the third row of the neutrino mixing matrix, and
solar sum rules can be derived from the conditions~\cite{Antusch:2007rk}
\beq \left | U_{\tau 1} \right |  = s^{\nu}_{12}s^{\nu}_{23}\; ,\ \ \ \ \left |
U_{\tau 2} \right |     = c^{\nu}_{12}s^{\nu}_{23} \; , \ \ \ \ \left | U_{\tau
3} \right |  = c^{\nu}_{23}  \;,  \label{sol} \eeq 
where $s^\alpha_{ij}\equiv \sin\theta^\alpha_{ij}$ and $c^\alpha_{ij}\equiv
\cos\theta^\alpha_{ij}$.
For example, with TBM neutrino mixing $s^{\nu}_{23}=c^{\nu}_{23}=1/\sqrt{2}$,
$s^{\nu}_{12}=1/\sqrt{3}$ and $c^{\nu}_{12}=\sqrt{2/3}$, 
\begin{align} \left | U_{\tau 1} \right |  = \frac{1}{\sqrt{6}}\; ,\ \ \ \
\left | U_{\tau 2} \right |     = \frac{1}{\sqrt{3}} \; ,\ \ \ \ \left |
U_{\tau 3} \right |  = \frac{1}{\sqrt{2}}  \;.  \label{sol2} \end{align}
The solar sum rule $s= r\cos\delta + \mathcal{O}(a^2,r^2,s^2)$ can be derived
from these conditions~\cite{King:2007pr}, where
$s\equiv\sqrt{3}\sin\theta_{12}-1$.  It is
clear that the conditions on $\left | U_{\tau 1} \right | $ and $\left |
U_{\tau 2} \right | $ in \eref{sol2} are identical to the corresponding
conditions for TM1 and TM2 mixing, respectively. However the
conditions on the other elements of the PMNS mixing matrix are different, so
the resulting atmospheric and solar sum rules will also be different.

In this paper, we extend the above derivation of the solar sum rule to the more
general case where not only $\theta^e_{12},\theta^{\nu}_{12},\theta^{\nu}_{23}$
are non-zero but also $\theta^e_{23}$ is allowed to be non-zero and all complex
phases are kept arbitrary, while still keeping
$\theta^e_{13}=\theta^{\nu}_{13}=0$. As a result we shall find the remarkable
condition, 
\beq \frac{\left | U_{\tau 1} \right |}{\left | U_{\tau 2} \right |
}=\frac{s^{\nu}_{12}}{c^{\nu}_{12}}=t^{\nu}_{12}\;. \label{sol3} \eeq 
Of course, the condition in \eref{sol3} can be trivially derived from
\eref{sol}, assuming $\theta^e_{23}=0$. The notable feature is that \eref{sol3}
also holds independently of $\theta^e_{23}$ and of all complex phases. However,
\eref{sol} involves further relations which only hold for $\theta^e_{23}=0$.
These further relations can be used to eliminate the atmospheric angle,
providing a solar sum rule which is more restrictive than that coming from
\eref{sol3} alone. Nevertheless, we shall continue to refer to the relation in
\eref{sol3} as a solar sum rule, since it is satisfied even when
$\theta^e_{23}=0$, as in \eref{sol}, and is distinct from the atmospheric sum
rules discussed earlier.  The solar sum rule in \eref{sol3} may be cast as a
prediction for $\cos \delta$, as a function of the measured mixing angles and
$\theta^{\nu}_{12}$, 
\be \cos \delta =\frac
{t_{23}s^2_{12}+s^2_{13}c^2_{12}/t_{23}-s^{\nu2}_{12}(t_{23}+s^2_{13}/t_{23})}
{\sin 2\theta_{12}s_{13}},  \label{sol4} \ee 
an expression which had been derived previously using an alternative argument
in Refs.~{\protect\cite{Marzocca:2013cr}} and {\protect\cite{Petcov:2014laa}}.
We also highlight a second remarkable feature of \eref{sol4}, namely that it is
not only independent of $\theta^e_{23}$ but also of $\theta^{\nu}_{23}$.  The
sum rule in \eref{sol4} is specified by only fixing the value of
$s^{\nu}_{12}$. 
Therefore, we can enumerate the viable models of this type by deriving the
values of $\theta^\nu_{12}$ associated with those leading-order mixing patterns
with $\theta^\nu_{13}=0$ which are derivable from considerations of symmetry.
In this article, we shall show that this leads us to four well-motivated solar
sum rules: one based on \TBM mixing \cite{Harrison:2002er} where
$s^{\nu}_{12}=1/\sqrt{3}$, one based on bimaximal (\BM) mixing
\cite{Barger:1998ta} where $s^{\nu}_{12}=1/\sqrt{2}$ and two patterns based on
versions of golden ratio mixing including \GR with $t^{\nu}_{12}=1/\varphi$
\cite{Datta:2003qg} and \GRnew with $c^{\nu}_{12}=\varphi
/\sqrt{3}$~\cite{Lam:2011ag,deAdelhartToorop:2011re}, where
$\varphi=\frac{1+\sqrt{5}}{2}$ is the golden ratio. We shall also discuss the
viability of two leading-order patterns which have been invoked in the
literature called \DGR with $\theta^{\nu}_{12}=\pi
/5$~\cite{Rodejohann:2008ir}, and hexagonal (\HEX) mixing with
$\theta^{\nu}_{12}=\pi /6$~\cite{Albright:2010ap}. 

For each viable prediction we perform a study of the scope to test the sum rule
in \eref{sol4} within the current experimental programme. Over the next few
decades, significant new information will be provided on the leptonic mixing
matrix from two main sources: the next generation of medium-baseline reactor
(MR) experiments and long-baseline wide-band superbeams (WBB). The MR programme
primarily seeks to measure the interference between atmospheric and solar
neutrino oscillations at baseline distances of around $50$--$60$~km.  These
facilities have been shown to be sensitive to the mass hierarchy
{\protect\cite{Petcov:2001sy,Qian:2012xh,Ge:2012wj,Li:2013zyd}}.  There are two
main experiments working towards a MR facility, both building on successful
measurements of $\theta_{13}$ at a shorter baseline: JUNO
{\protect\cite{Wang:2013aa}} and RENO-50
{\protect\cite{Park:2014jda,Park:2014sja}}. The WBB experiments can be seen as
complementary to the MR proposals. Collaborations such as LBNE
{\protect\cite{Adams:2013qkq}} and LBNO {\protect\cite{Agarwalla:2013kaa}}
intend to construct a high-power long-baseline neutrino and antineutrino beam
which can exploit matter effects and large statistics for the primary aim of
constraining the CPV phase $\delta$. The combination of MR and WBB facilities
will provide new levels of precision in the neutrino sector, with $\theta_{12}$
and $\theta_{13}$ being probed to the level of percent by MR experiments, and
$\delta$ being constrained by dedicated WBB facilties. This complementarity
offers for the first time the possibility of experimentally testing relations
such as {\protect\eref{sol4}}.\footnote{For another application of this
complementarity to the study of flavour-symmetric predictions, see
\refref{Ballett:2014uia}} In this work, we shall simulate illustrative MR and
WBB facilities with an aim to exploring how their complementarity can be used
to constrain the models of charged-lepton corrections.

The idea of correcting a leading-order neutrino mixing pattern by
contributions from the charged leptons has recently been revisited
\cite{Marzocca:2013cr,Petcov:2014laa,Gollu:2013yla,Sruthilaya:2014kca,Girardi:2014faa}. Our work
goes beyond these analyses in three ways. Firstly, we present a novel
derivation of the correlation in {\protect\eref{sol4}} in a more general setting, showing
it to be the consequence of the simpler relation {\protect\eref{sol3}}, which helps to
highlight its relationship to the earlier solar sum rules. Secondly, we
systematically derive the possible leading-order mixing patterns from
considerations of residual symmetry, finding a small well-motivated set.
Thirdly, we present the results of simulations assessing the potential to
constrain the solar sum rules from two upcoming complementary oscillation
experiments: a superbeam and a reactor facility.

The layout of the remainder of the paper is as follows: in
\secref{sec:derivation}, we present a simple derivation of the correlations
induced by charged-lepton corrections. We then systematically identify the
viable leading-order neutrino mixing matrices, and comment on their relation to
the underlying flavour symmetry.  \secref{sec:numerics} is devoted to our
numerical study. We first consider the currently allowed parameter spaces of
these correlations, then we present the details and results of our simulations
of a superbeam and reactor experiment, showing how these can be used to test
these relations.  We comment on the case where $\theta^\nu_{13} \neq 0$ in
\secref{sec:beyond}, and discuss renormalisation group effects in
\secref{sec:rge}.  Finally, \secref{sec:conclusion} concludes the paper.

%%%%%%%%%%%%%%%%%%%%%%%%%%%%%%%%%%%%%%%%%%%%%%%%%%%%%%%%%%%%%%%%%%%%%%
\section{\label{sec:derivation}Mixing sum rules from charged-lepton
corrections} In the first subsection, we present a simple derivation of the
solar sum rule of \eref{sol4}. Then in later subsections we discuss the
leading-order mixing patterns which one encounters in the considered class of
models. We shall find that there are only four well-motivated patterns of
interest, whose relation to model building will be discussed.

\subsection{A simple derivation}
In the equations that follow, superscripts are attached to quantities which are
naturally associated with the neutrinos or the charged leptons (\eg
$\theta^\nu$ and $\theta^e$), whilst physical parameters go without.

Assuming\footnote{It is possible to derive sum rules with $\theta_{13}^\nu\neq
0$. We comment on one example in \secref{sec:beyond}.}
$\theta^{\nu}_{13}=\theta^e_{13}=0$, the PMNS matrix is given by the product
of five unitary matrices  
\begin{equation} 
\label{123}
U=U^{e\dagger}_{12}U^{e\dagger}_{23}R^{\nu}_{23}R^{\nu}_{12}P^{\nu},
\end{equation}
the three right-most matrices describe the neutrino sector, and are
parameterised by  
\begin{align*} 
R^{\nu}_{23}= \left(\begin{matrix} 1 & 0 & 0 \\ 0 & c^{\nu}_{23} & s^{\nu}_{23}
\\ 0 & -s^{\nu}_{23} & c^{\nu}_{23}\end{matrix}\right)\quad\text{and}\quad~~
R^{\nu}_{12}= \left(\begin{matrix} c^{\nu}_{12} & s^{\nu}_{12} & 0 \\
-s^{\nu}_{12}& c^{\nu}_{12} & 0\\ 0 & 0 & 1 \end{matrix}\right), \end{align*} 
and $P^\nu$ is a diagonal matrix of
uni-modular complex numbers. The two unitary matrices on the left of \eref{123}
characterise the charged-lepton corrections, and will be allowed to include
extra complex phases, 
\begin{align*} U^e_{23}& = \left(\begin{array}{ccc} 1 & 0 & 0 \\ 0 & c^e_{23} &
s^e_{23}e^{-i\delta^e_{23}} \\ 0 & -s^e_{23}e^{i\delta^e_{23}} & c^e_{23} \\
\end{array}\right),\\ U^e_{12}&= \left(\begin{array}{ccc} c^e_{12} &
s^e_{12}e^{-i\delta^e_{12}} & 0 \\ -s^e_{12}e^{i\delta^e_{12}} & c^e_{12} & 0\\
0 & 0 & 1  \end{array}\right).  \end{align*} 

With these definitions, it is simple enough to compute the explicit form of the
PMNS matrix. However, our derivation focuses only on the first two elements of
the bottom row of the physical PMNS matrix, which are found to be 
\begin{equation}
\begin{aligned} U_{\tau1} &=
s^{\nu}_{12}(s^{\nu}_{23}c^e_{23}-c^{\nu}_{23}s^e_{23}e^{i\delta^e_{23}} ),\\
U_{\tau2} &=
-c^{\nu}_{12}(s^{\nu}_{23}c^e_{23}-c^{\nu}_{23}s^e_{23}e^{i\delta^e_{23}}).
\end{aligned} \label{Ucorr}\end{equation}
By comparing \eref{Ucorr} to the PDG parameterisation of $U$
\cite{Agashe:2014kda}, we find the relations between the physical parameters
and our internal parameters, 
\begin{align*} 
|U_{\tau1}| = |s_{23}s_{12}-s_{13}c_{23}c_{12}e^{i\delta} |    &=
|s^{\nu}_{12}(s^{\nu}_{23}c^e_{23}-c^{\nu}_{23}s^e_{23}e^{i\delta^e_{23}})|\;,\\
|U_{\tau2}| = | s_{23}c_{12}+s_{13}c_{23}s_{12}e^{i\delta} |    &=
|c^{\nu}_{12}(s^{\nu}_{23}c^e_{23}-c^{\nu}_{23}s^e_{23}e^{i\delta^e_{23}} )|\;.
\end{align*} 
As the ratio of these two equations is independent of the values of the parameters
in $U^e_{23}$ and $U^e_{12}$, we are left with a correlation between
observable parameters and the value of the neutrino mixing parameter
$\theta^\nu_{12}$,  
\be 
\frac{\left | U_{\tau 1} \right |}{\left | U_{\tau 2} \right |}=
\frac{ |s_{23}s_{12}-s_{13}c_{23}c_{12}e^{i\delta} | } {  |
s_{23}c_{12}+s_{13}c_{23}s_{12}e^{i\delta} |  }=t^{\nu}_{12}.  \label{sum1} \ee
This correlation will be referred to as the solar mixing sum rule. It can be
viewed as a predictive statement about the physical CPV phase: squaring both
sides of \eref{sum1} and solving for $\cos \delta$ leads us to the expression
in \eref{sol4}, which we repeat below,
\be \cos \delta =\frac
{t_{23}s^2_{12}+s^2_{13}c^2_{12}/t_{23}-s^{\nu2}_{12}(t_{23}+s^2_{13}/t_{23})}
{\sin 2\theta_{12}s_{13}}.  \label{sol4p} \ee 

An equivalent correlation has been derived previously using a lengthier argument
in Refs.~\cite{Marzocca:2013cr} and \cite{Petcov:2014laa}. Understanding its application to specific
models, its compatibility with global data and its potential use as a signature
of new physics will be the focus of the rest of this article.

\begin{figure}[t] \begin{tabular}{c c}
\includegraphics[width=0.49\linewidth]{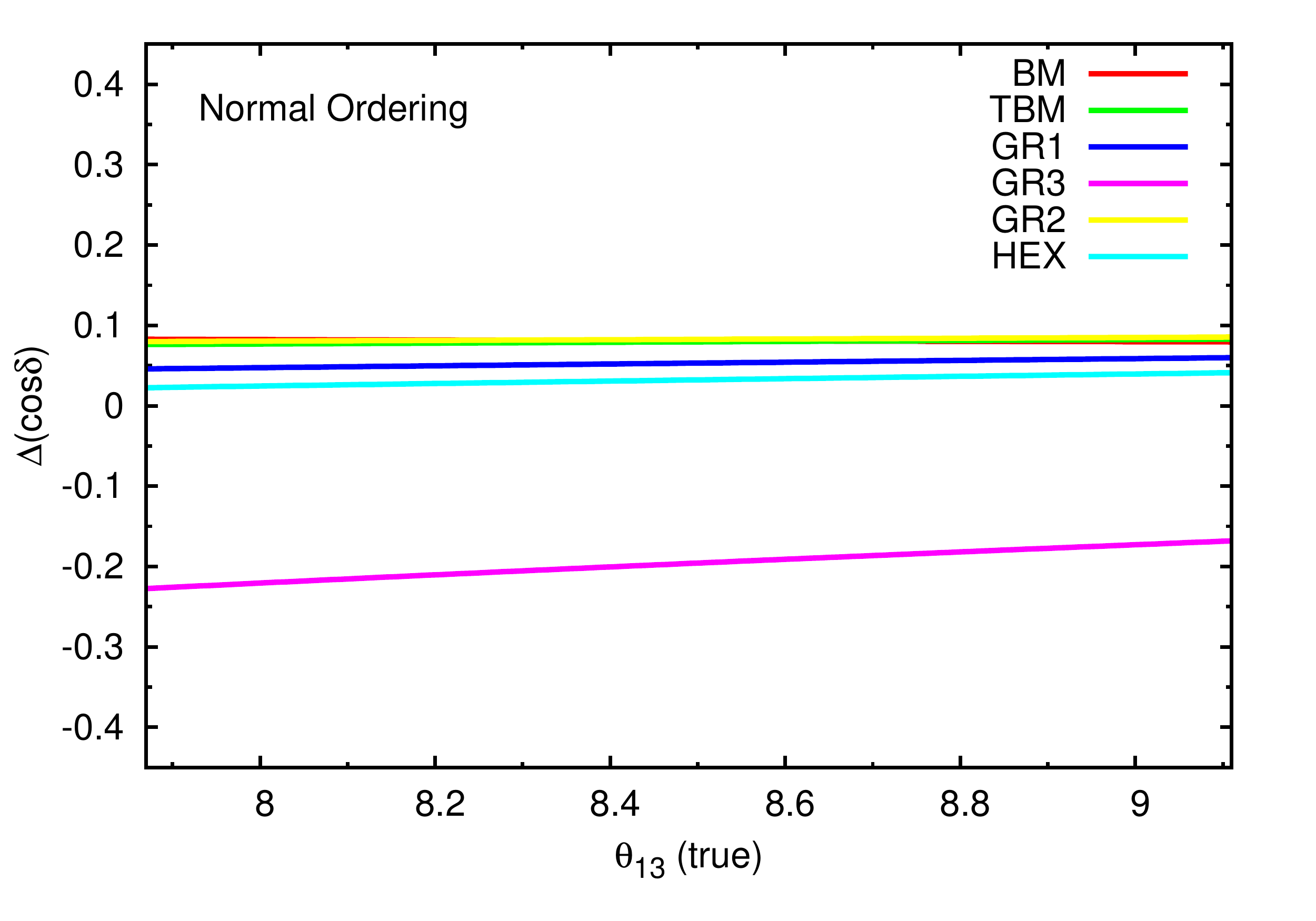} &
\includegraphics[width=0.49\linewidth]{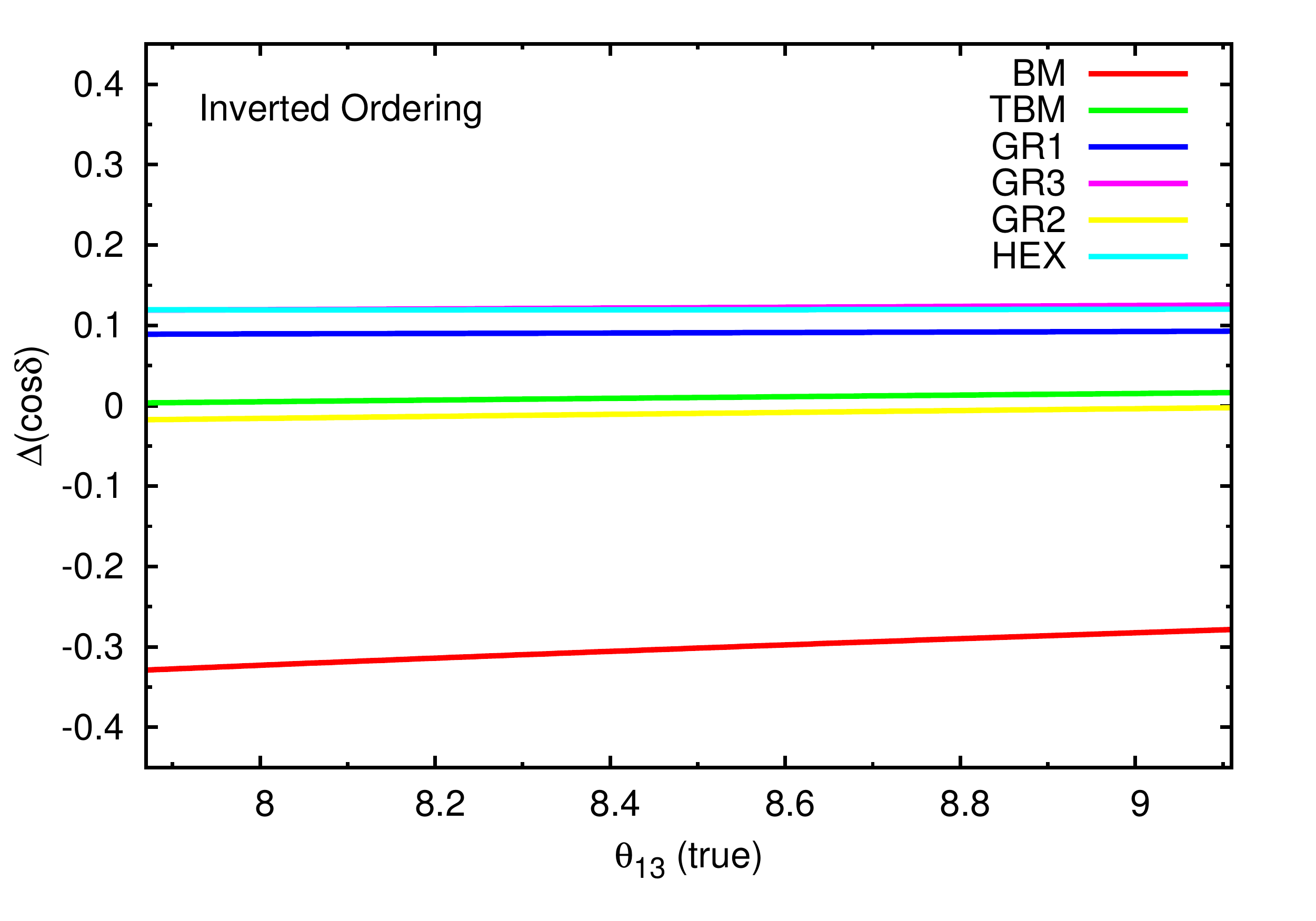} \end{tabular}
\caption{\label{fig:cosdelta_error}The difference between the
linearised expression
$\cos\delta_\text{linear}\equiv(\theta_{12}-\theta_{12}^\nu)/\theta_{13}$ and the solar sum rule
in {\protect\eref{sol4}}. These plots assume $\theta_{12} = 33.5^\circ$ and
take $\theta_{23}$ to be the best-fit value for normal (inverted)
ordering in the left (right) panel.
The best-fit values are those of \refref{Gonzalez-Garcia:2014bfa}.}
\end{figure}

The correlation in \eref{sol4} is in fact the full non-linear version of a more
familiar first-order relation. We collect a number of phenomenologically
interesting approximations in \appref{sec:simpApp}. 
If we expand \eref{sol4} in a small parameter $\varepsilon$, assumed to control
the deviation from a leading-order neutrino mixing pattern with maximal
atmospheric mixing, 
\begin{equation}  \theta_{13} ~\sim ~ \left | \theta_{12} - \theta^\nu_{12}
\right | ~\sim~ \left |\theta_{23}-\frac{\pi}{4} \right | ~\sim ~ \varepsilon,
\label{eq:approx_parameters_linear}\end{equation}
we find the well-known first-order relation~
\cite{King:2005bj,Masina:2005hf,Antusch:2005kw},
\begin{equation} \theta_{12}= \theta^{\nu}_{12} + \theta_{13}\cos\delta +
\mathcal{O}(\varepsilon^2). \label{eq:linearSSR} \end{equation}
The validity of this approximation is dependent upon the severity of the
assumptions in \eref{eq:approx_parameters_linear}. This can only be assessed on
a model dependent basis; however, in \figref{fig:cosdelta_error} we show the
size of the error $\Delta(\cos\delta) \equiv \cos\delta_\text{linear} -
\cos\delta$ which is introduced by the linear approximation for the patterns
which we will derive in subsections {\protect\ref{sec:fully}} and
{\protect\ref{sec:partially}}. Apart from the patterns denoted {\protect\GRnew}
and {\protect\BM} (which we will argue in the following section are strongly
disfavoured by current data), the error approximately satisfies
$|\Delta(\cos\delta)|\lesssim 0.1$.
As deviations of this size are expected to be close to the attainable precision
at a next-generation oscillation facility, all subsequent numerical work will
use the full correlations in \eref{sol4}.\footnote{It has been argued
{\protect\cite{Girardi:2014faa,Petcov:2014laa}} that the ratio of leading-order
to exact predictions indicate that the linearized sum rules are not accurate
enough for phenomenological use. We believe that for many purposes the
linearized expressions would be adequate: constant errors of
$\Delta(\cos\delta)=0.1$ induce an error of less than $15^\circ$ ($10^\circ$)
for $76\%$ ($60\%$) of the range of $\delta$. Therefore the linearized
expressions well describe the correlation to the precision of the first phases
of the next-generation of superbeams, which expect a sensitivity of
$15$--$30^\circ$ {\protect{\cite{Coloma:2012wq,deGouvea:2013onf}}}; however,
the full expressions will be necessary in the subsequent phases, where
precisions are expected to be $8$--$18^\circ$
{\protect{\cite{Coloma:2012wq,deGouvea:2013onf}}}.}

The solar sum rule derived above is valid for any neutrino mixing pattern with
$\theta^{\nu}_{13}=0$ and for any charged-lepton corrections with
$\theta^e_{13}=0$.
Our focus in this work is on the predictions of models which apply
charged-lepton corrections to neutrino mixing matrices which are completely
fixed by symmetry. In recent work, significant progress has been made in the
categorisation of fully-specified mixing patterns subject to some weak model
building assumptions. In \subsecref{sec:fully}, we shall identify a set
of leading-order predictions with $\theta_{13}^\nu=0$ from arguments of
symmetry by following two categorisation schemes from the
literature~\cite{Hernandez:2012ra,Fonseca:2014koa}. As we 
shall explain, strictly speaking, one of these
frameworks~\cite{Hernandez:2012ra} is a subcase of the
other~\cite{Fonseca:2014koa}; however, its systematic exploration has not been
presented before, and we shall show how this more restrictive scenario still
finds all of the cases of the more comprehensive analysis, while shedding
light on the group structure of the viable solutions.

In \subsecref{sec:partially}, we shall also comment on some mixing patterns
frequently invoked in the literature which are not found in the systematic
derivations. We will discuss these patterns in the context of an infinite
family of neutrino mixing matrices which are partially constrained by symmetry. For the
lack of a symmetric origin, we believe these patterns to be more poorly
motivated; however, we shall include them in our numerical analysis for
completeness.

\subsection{\label{sec:fully}Fully specified mixing patterns}

An impressively comprehensive account of fully specified leading-order
mixing matrices has been presented in \refref{Fonseca:2014koa}. In
this work, it was assumed that neutrinos are Majorana particles, that a finite
flavour group $G$ is broken into the Klein group $G_\nu
= \mathbb{Z}_2 \times \mathbb{Z}_2$ in the neutrino sector, and $G_e = \mathbb
Z_n$, with $n\in \mathbb N$, in the charged-lepton sector. Such an arrangement
completely specifies the leading-order mixing matrix. Under these  
general assumptions, it was shown that the only possible mixing matrices are
given by 17 sporadic patterns and one infinite family of patterns (up to row and
column permutations). In \refref{Fonseca:2014koa}, all $17$ sporadic patterns
are shown to be excluded by the current global neutrino oscillation data at
$3\sigma$, whilst the infinite family is allowed for some values of its
parameters. In the current work, we are expecting corrections to the
leading-order mixing angles of a magnitude $\theta_{13}$, and therefore we
have rather more lenient criteria for viability. We define the eligible
leading-order mixing patterns as those which meet the criteria 
$\theta_{13}^\nu\le20^\circ$,
$20^\circ\le\theta_{12}^\nu\le45^\circ$ and 
$30^\circ\le\theta_{23}^\nu\le60^\circ$. 
Scanning over the patterns found in \refref{Fonseca:2014koa} (including row and
column permutations), we find that the 17 sporadic patterns allow 13 viable
matrices. The infinite family meets our criteria for about $20\%$ of its
allowed parameter space.
However, our present aim is to discuss situations where a leading-order pattern
with $\theta^\nu_{13}=0$ can be brought in-line with observation through
corrections from the charged-lepton sector. If we therefore restrict our attention to
patterns with $\theta_{13}^\nu=0$, we find only $4$ patterns 
which pass our lax phenomenological conditions on the remaining two neutrino
mixing angles $\theta_{12}^\nu$ and $\theta_{23}^\nu$. It is interesting to
note that all 4 patterns differ from one another only in their value for the
solar mixing angle $\theta_{12}^\nu$ with the atmospheric mixing angle fixed
at $\theta_{23}^\nu=45^\circ$.

The first eligible pattern is known as bimaximal (\BM) mixing. It has a
maximal solar mixing angle \cite{Barger:1998ta}, and is given by a matrix of
the form 
\begin{equation}\tag{BM}
U^\nu_\text{BM} = \left(\begin{matrix} \frac{1}{\sqrt{2}} &  \frac{1}{\sqrt{2}} &
0 \\ - \frac{1}{2}  & \frac{1}{2} &  \frac{1}{\sqrt{2}} \\ \frac{1}{2} &
-\frac{1}{2} &  \frac{1}{\sqrt{2}}    \end{matrix} \right). 
\end{equation}
The second pattern is the tribimaximal (\TBM) mixing matrix. This has been
associated with models based on the flavour symmetries A$_4$ and S$_4$. It
predicts a solar mixing angle given by $s^{\nu}_{12}=1/\sqrt{3}$,
i.e. $\theta_{12}^\nu\approx 35.3^\circ$. The mixing matrix is given explicitly by
\begin{equation}\tag{TBM}
U^\nu_\text{TBM} = \left(\begin{matrix} \sqrt{\frac{2}{3}} &  \frac{1}{\sqrt{3}}
&  0 \\ - \frac{1}{\sqrt{6}}  & \frac{1}{\sqrt{3}} &  \frac{1}{\sqrt{2}} \\
\frac{1}{\sqrt{6}} & -\frac{1}{\sqrt{3}} &  \frac{1}{\sqrt{2}}    \end{matrix}
\right).
\end{equation}
The remaining two patterns both associate the golden ratio
$\varphi=\frac{1+\sqrt{5}}{2}$ with the solar mixing angle, although in
different ways. The first is the original golden ratio mixing pattern
(\GR)~\cite{Datta:2003qg}, related to the flavour symmetry $A_5$. It predicts
$t_{12}^\nu=1/\varphi$, i.e. $\theta^\nu_{12}\approx 31.7^\circ$, resulting in
the mixing matrix
\begin{equation}\tag{\GR}
 U^\nu_\text{\GR} = \left( \begin{matrix} \frac{\varphi}{\sqrt{2+\varphi}} &
\frac{1}{\sqrt{2+\varphi}} &  0 \\ -\frac{1}{\sqrt{4+2\varphi}}&
\frac{\varphi}{\sqrt{4+2\varphi}} & \frac{1}{\sqrt{2}} \\
\frac{1}{\sqrt{4+2\varphi}} & -\frac{\varphi}{\sqrt{4+2\varphi}} &
\frac{1}{\sqrt{2}} \end{matrix}\right).
\end{equation}
The other golden ratio pattern found by our survey is less well known but has
been found previously in \refsref{Lam:2011ag}{deAdelhartToorop:2011re}. It is
associated with the group A$_5$ breaking into a $\mathbb{Z}_3$ symmetry in the
charged-lepton sector and the Klein symmetry in the neutrino sector. This
pattern (\GRnew) predicts $c_{12}^\nu = \varphi/\sqrt{3}$,
i.e. $\theta^\nu_{12}\approx 20.9^\circ$, leading to a mixing matrix of the
form
\begin{equation}
\tag{\GRnew}
U^\nu_\text{\GRnew} = \left( \begin{matrix} \frac{\varphi}{\sqrt{3}} &
\frac{\gal{\varphi}}{\sqrt{3}}  &  0 \\ -\frac{\gal{\varphi}}{\sqrt{6}} &
\frac{\varphi}{\sqrt{6}}  & \frac{1}{\sqrt{2}} \\
\frac{\gal{\varphi}}{\sqrt{6}} & -\frac{\varphi}{\sqrt{6}} & \frac{1}{\sqrt{2}}
\end{matrix}\right), 
\end{equation}
where $\gal{\varphi}$ is the Galois ($\mathbb{Q}$-)conjugate of $\varphi$ given
by $\gal{\varphi}=\frac{1-\sqrt{5}}{2}$. 
To make the connection with the nomenclature of Fonseca and Grimus
\cite{Fonseca:2014koa}: \BM is known as $\mathcal{C}_1$, \TBM is the only
member of the infinite family $\mathcal{C}_2$ with $\theta_{13}^\nu=0$, \GR is
$\mathcal{C}_{11}$ and \GRnew is known as $\mathcal{C}_{12}$.

We will now show that these four patterns can also be derived under the
framework of Refs.~\cite{Hernandez:2012ra,Hernandez:2012sk} in 
which the assumption of $\theta^\nu_{13}=0$ will be shown to be unnecessary.  
This scenario can be seen as a subcase of the previous systematic analysis, also 
working under the assumption of Majorana neutrinos and a finite group $G$ broken 
into distinct residual symmetries amongst the charged-lepton sector and the neutrino sector.
However, a further assumption is made on the form of the finite groups: they
are assumed to be overgroups of the von Dyck groups
\cite{Hernandez:2012ra,Hernandez:2012sk}, 
\[ D(2,m,p) = \langle S,T,W | S^2 = T^m = W^p = STW = 1 \rangle.  \]
Finiteness of $D(2,m,p)$ restricts the values of $\{m,p \}$ to either
$\{3,3\}$, $\{3,4\}$, $\{3,5\}$ or $\{2,N\}$, where the first three choices are
associated with the groups A$_4$, S$_4$ and A$_5$, respectively, while the
fourth is related to the dihedral groups $D_{2N}$.  The fully specified mixing
matrices were not systematically derived in \refref{Hernandez:2012sk}, and we
will now sketch this calculation, deferring details to \appref{app:klein}. It
is particularly interesting to note that the further constraint on the form of
the group in this framework makes the restriction $\theta^\nu_{13}=0$
unnecessary: the only patterns meeting our phenomenological selection criteria
are the four previous patterns with $\theta^\nu_{13}=0$. 

Taking the generators of the symmetry of the charged lepton mass terms to be
$T$ and those of the Klein group acting on the neutrino mass terms as $S_1$ and
$S_2$, constraints can be derived on the leading-order mixing matrix
directly. Disregarding 
cases related to the dihedral symmetries,\footnote{We comment on the option
where $\{m,p \}=\{2,N\}$ at the end of \appref{app:klein}.} we find that the patterns are
specified by a choice of two parameters $\eta_1$ and $\eta_2$ taken from the
set 
\begin{equation}\left\{\frac{1}{3}, \frac{2}{3}, \frac{1}{2}, \frac{1+\varphi}{3},
\frac{2-\varphi}{3},\frac{2+\varphi}{5}, \frac{3-\varphi}{5} \right \}, 
\label{eq:choices}
\end{equation}
subject to the unitarity constraint $\eta_1 + \eta_2\le1$. The squared moduli of the
elements of the mixing matrix are then given by the following pattern (up to row
and column permutations) 
\begin{equation}  |U^\nu_{\alpha i}|^2 =  \left( \begin{matrix} \eta_1 & \eta_2 &
1-\eta_1-\eta_2 \\ \frac{1-\eta_1}{2} & \frac{1-\eta_2}{2} &
\frac{\eta_1+\eta_2}{2}\\ \frac{1-\eta_1}{2} & \frac{1-\eta_2}{2} &
\frac{\eta_1+\eta_2}{2} \end{matrix}\right).\end{equation}
Of all the possible combinations of $\eta_1$ and $\eta_2$,
only four leading-order neutrino mixing patterns are eligible by our criteria
on~$\theta^\nu_{ij}$ stated at the beginning of this subsection. They are
exactly those found in our discussion above:   
bimaximal mixing, tribimaximal mixing, and two patterns associated with
the golden ratio (\GR and \GRnew). 

\subsection{\label{sec:partially}Common partially constrained patterns}

There are a few other common mixing patterns with $\theta_{13}^\nu = 0$.
However, these patterns are not found in the systematic surveys of the previous
subsection. 
We will focus on two patterns of this type mentioned in the literature: one
associated with the golden ratio (\DGR)~\cite{Rodejohann:2008ir} and one
called hexagonal mixing (\HEX)~\cite{Albright:2010ap}. 
Both patterns have maximal atmospheric mixing $\theta^\nu_{23}=\pi/4$ and
vanishing reactor angle $\theta^\nu_{13}=0$, but they differ in their
predictions for $\theta^\nu_{12}$,
\begin{align*} \theta^\nu_{12} = \frac{\pi}{5}\quad\text{(\DGR)}\qquad\text{and}\qquad\theta^\nu_{12} = \frac{\pi}{6}\quad\text{(\HEX)}.  \end{align*}
These predictions can be understood as part of a family of patterns which
predict $\theta_{12}^\nu =\frac{\pi d}{N}$ with $d,N\in \mathbb N$ and $0< d < N$. 
These are commonly connected to the dihedral groups $D_{2N}$, and indeed it is
possible to derive partial constraints consistent with these patterns by
breaking $D_{2N}$ to different preserved subgroups in the charged-lepton
sector and the neutrino sector~\cite{Blum:2007jz}.
Such a construction can generate the prediction
\[\cos\theta^\nu_{12}\cos\theta^\nu_{13} = \cos\left(\frac{\pi d}{N}\right),\]
which leads to the patterns of interest if we fix $\theta_{13}^\nu=0$ by hand.
Furthermore, the assumption of a dihedral group as the fundamental flavour
symmetry does not permit the unification of the three families into a single
irreducible representation of the symmetry group.\footnote{All dihedral groups
have irreducible representations of dimensions $1$ and $2$ only.} 
For these reasons we consider these mixing patterns to be on less firm footing
than the models arising from the construction presented previously. Despite
these reservations, we will include the \DGR and \HEX neutrino mixing patterns
in our later analysis. However, we would like to point out that any rational
multiple of $\pi$ can be found for $\theta^\nu_{12}$ using a suitably large
dihedral group in this fashion; \DGR and \HEX are only distinguished in that
they are the best-fitting predictions of the form $\theta^\nu_{12}=\frac{\pi}{N}$.

\section{\label{sec:numerics}Numerical results}

\subsection{\label{sec:predictions}Allowed parameter spaces for solar sum rules}

The parameter correlations discussed above can be seen as predictions for the
remaining unknown parameter $\delta$. We will require $\cos \delta$ to lie in
the physical region, for which $-1 \leq \cos \delta \leq 1$. This may not occur
for all models considered or in all the allowed parameter space.  In this
section, we will consider the predictions of the solar sum rules over the
current $3\sigma$ interval of the mixing angles. We will use the values from v2
of the NuFit collaboration~\cite{Gonzalez-Garcia:2014bfa}. This parameter space
is denoted by $I_{3\sigma}$, 
\begin{align*} I_{3\sigma} &= I_{12}\times I_{13}\times I_{23}\\ &=
[31.29^\circ,35.91^\circ]\times[7.87^\circ,9.11^\circ]\times[38.3^\circ,53.3^\circ].
\end{align*}
\begin{figure*}[tbp]
\centering
\begin{tabular}{c c}
\includegraphics[width=0.48\linewidth]{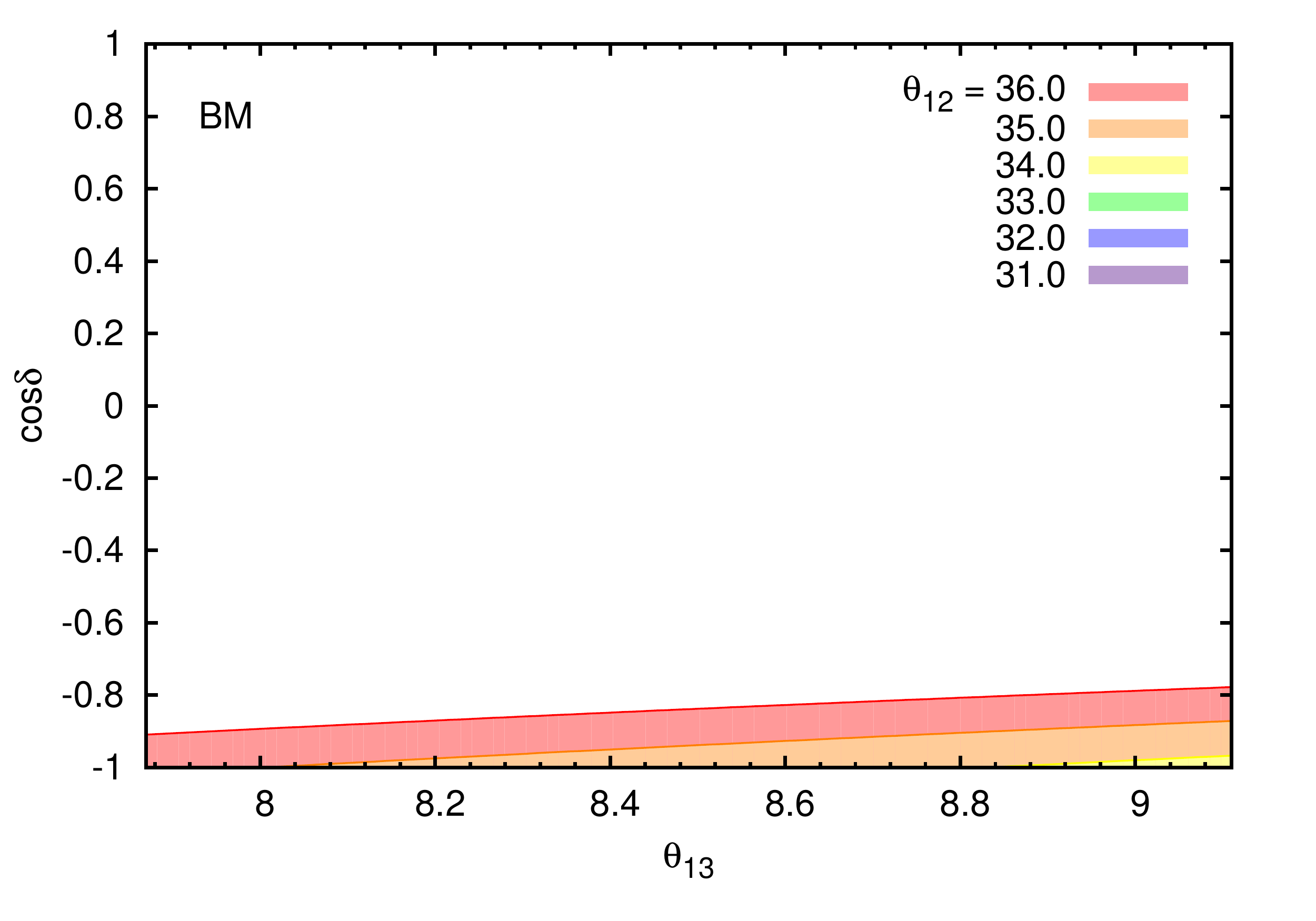} &
\includegraphics[width=0.48\linewidth]{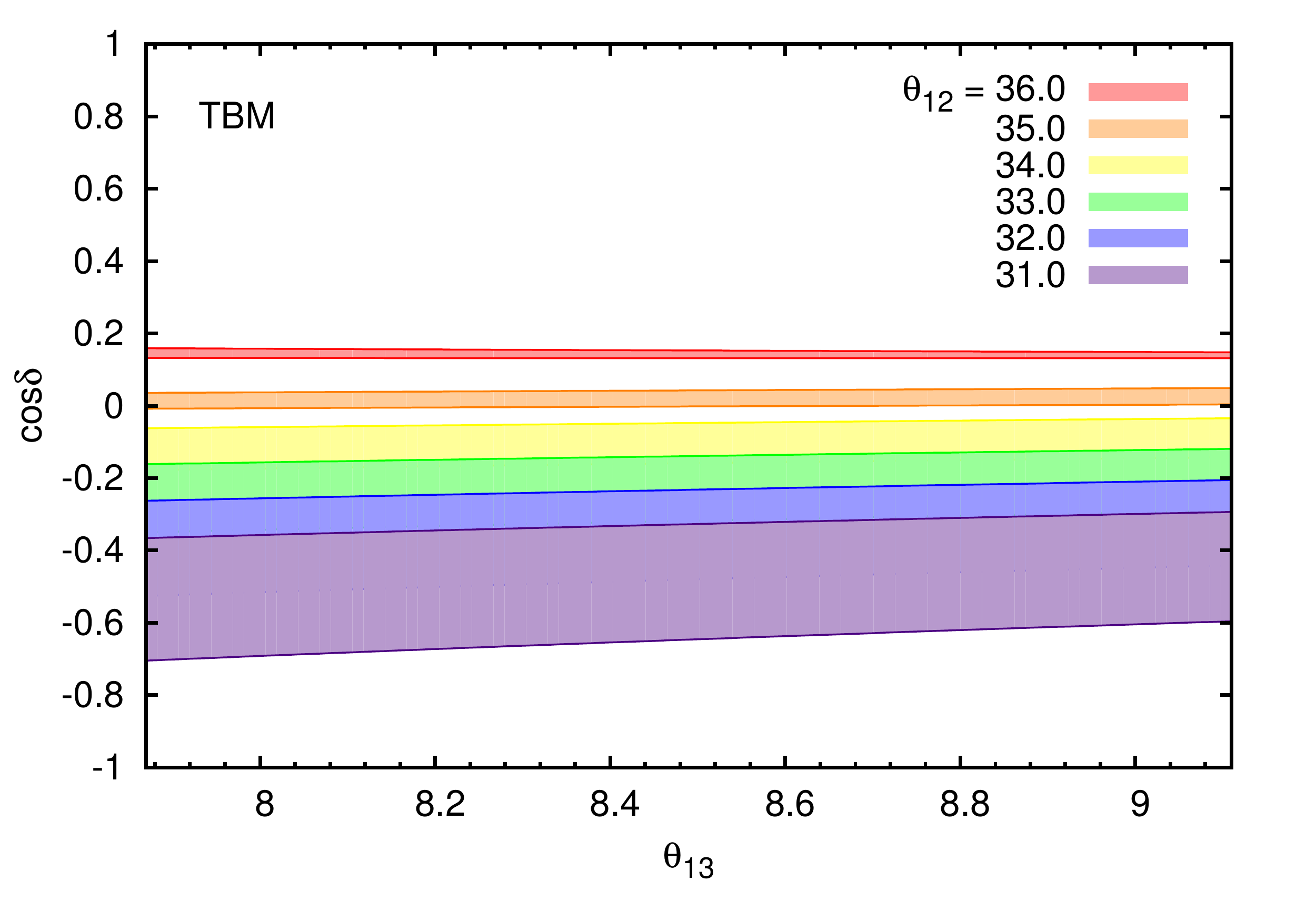} \\
\includegraphics[width=0.48\linewidth]{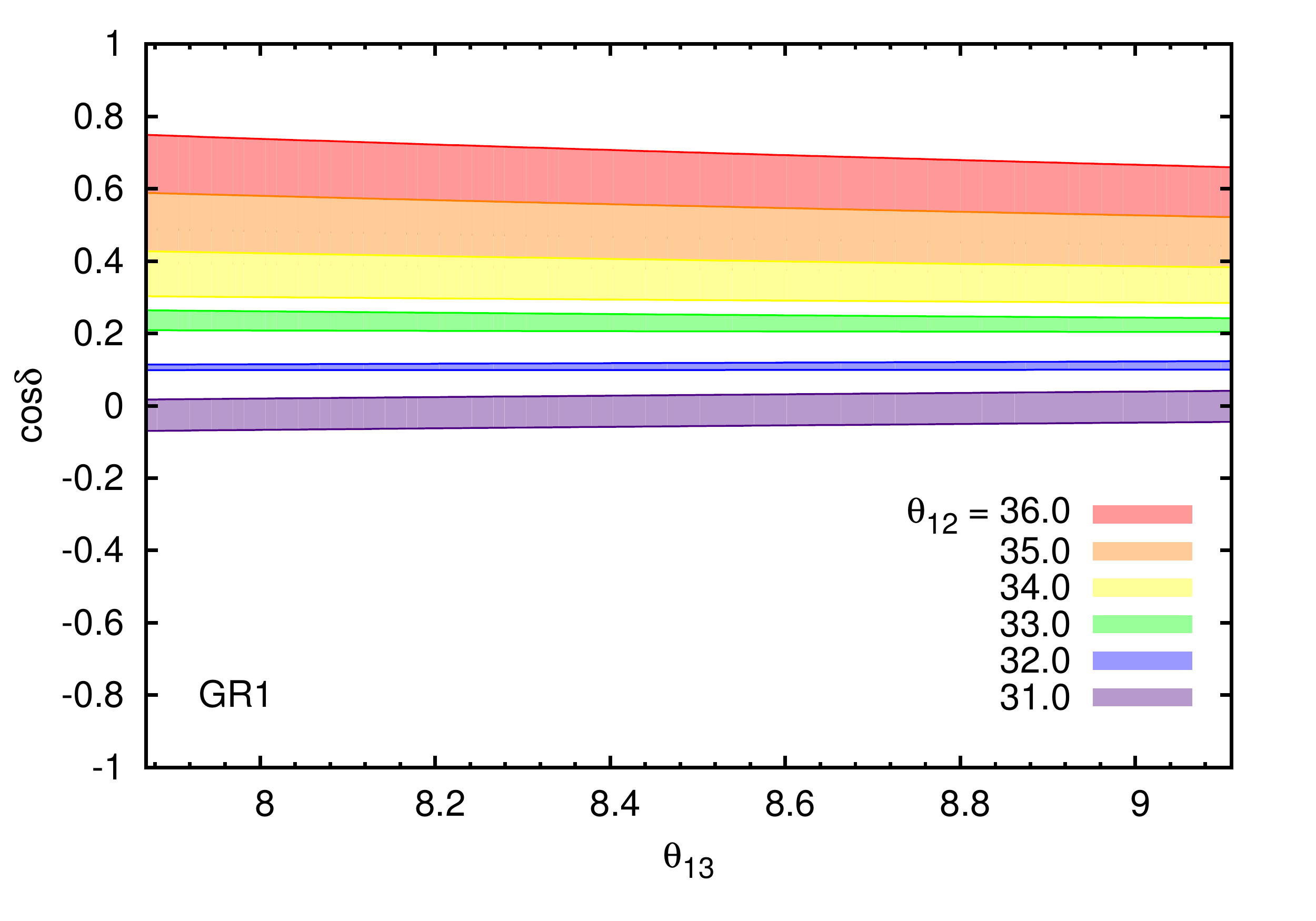} &
\includegraphics[width=0.48\linewidth]{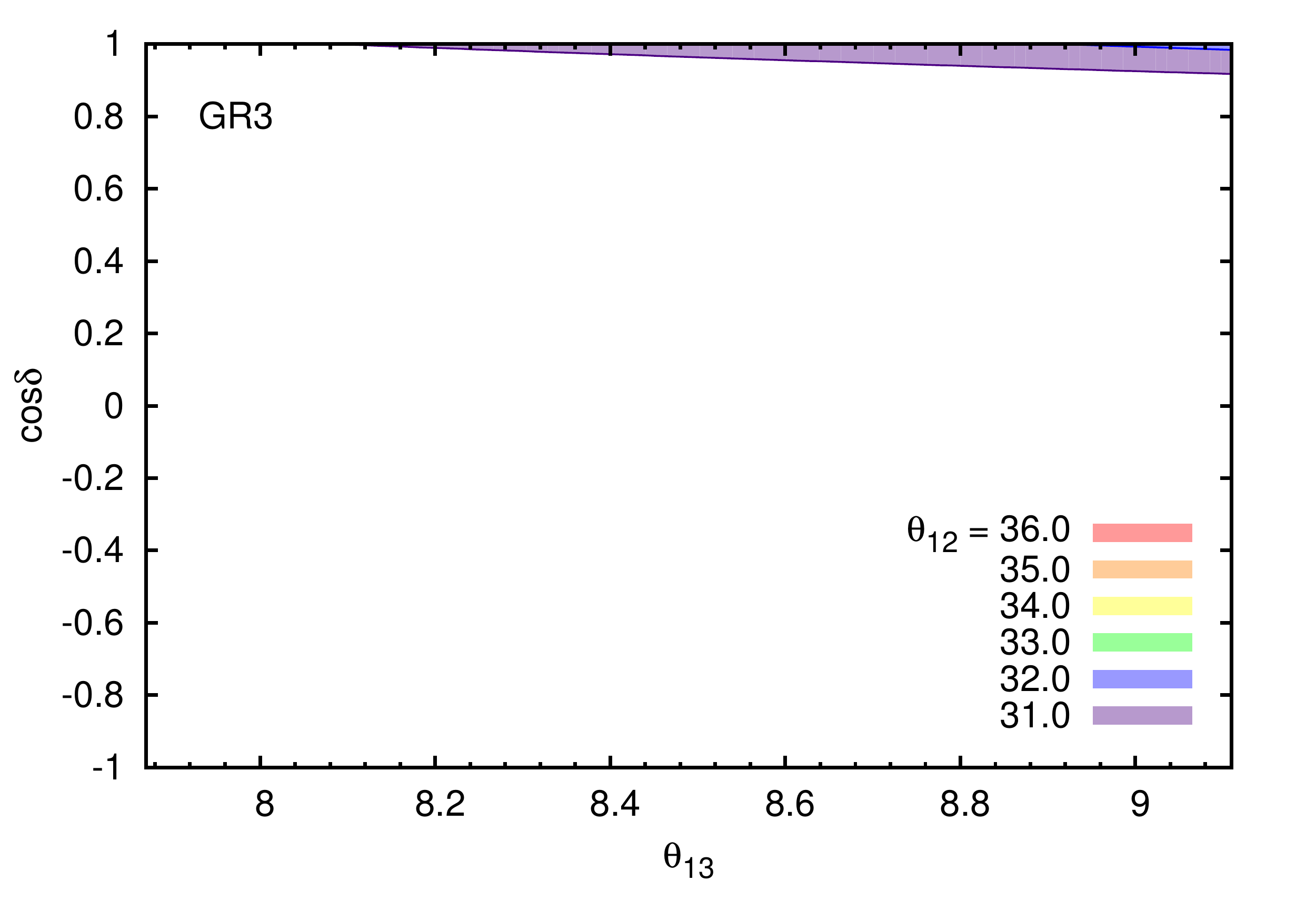} \\
\includegraphics[width=0.48\linewidth]{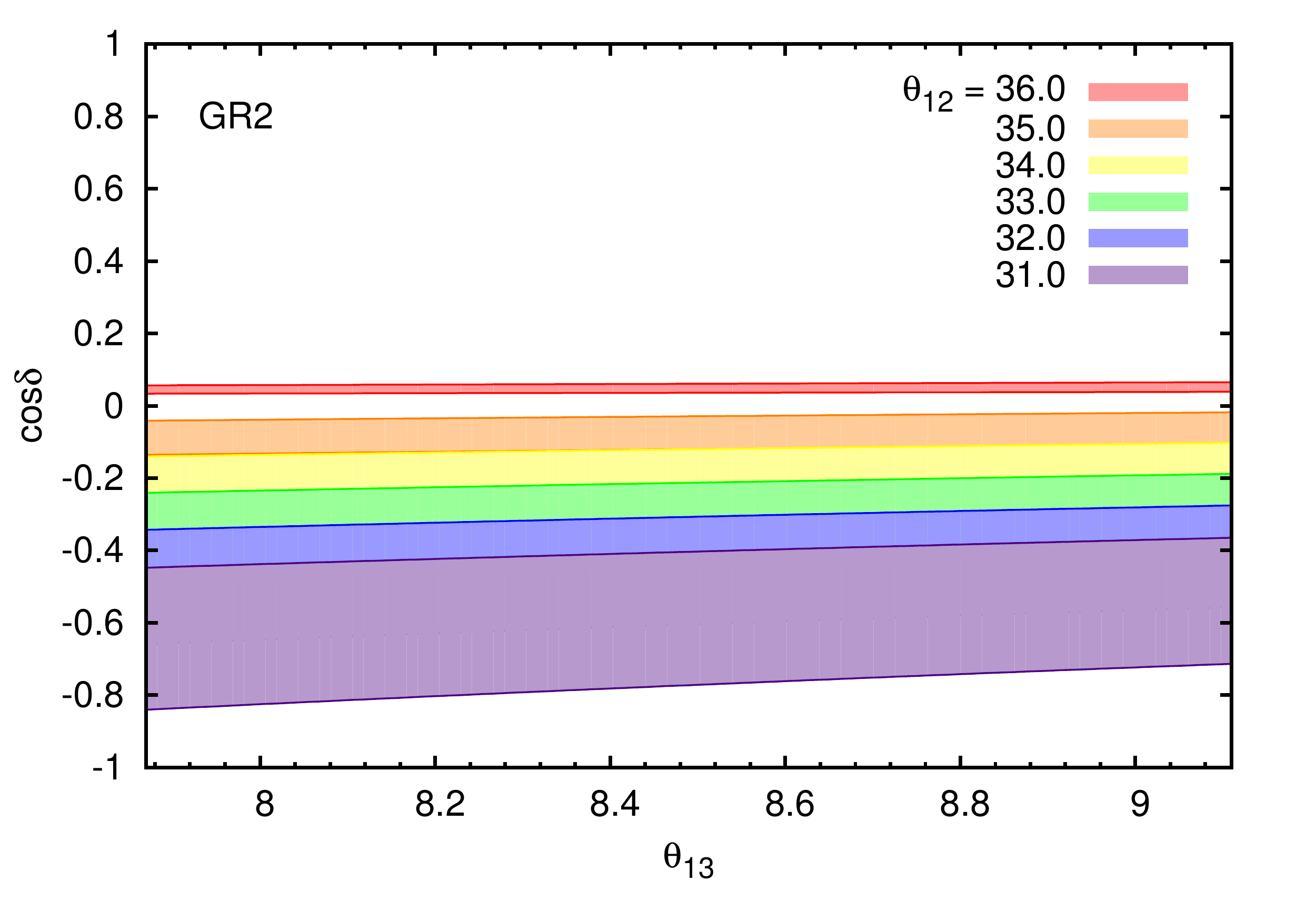} &
\includegraphics[width=0.48\linewidth]{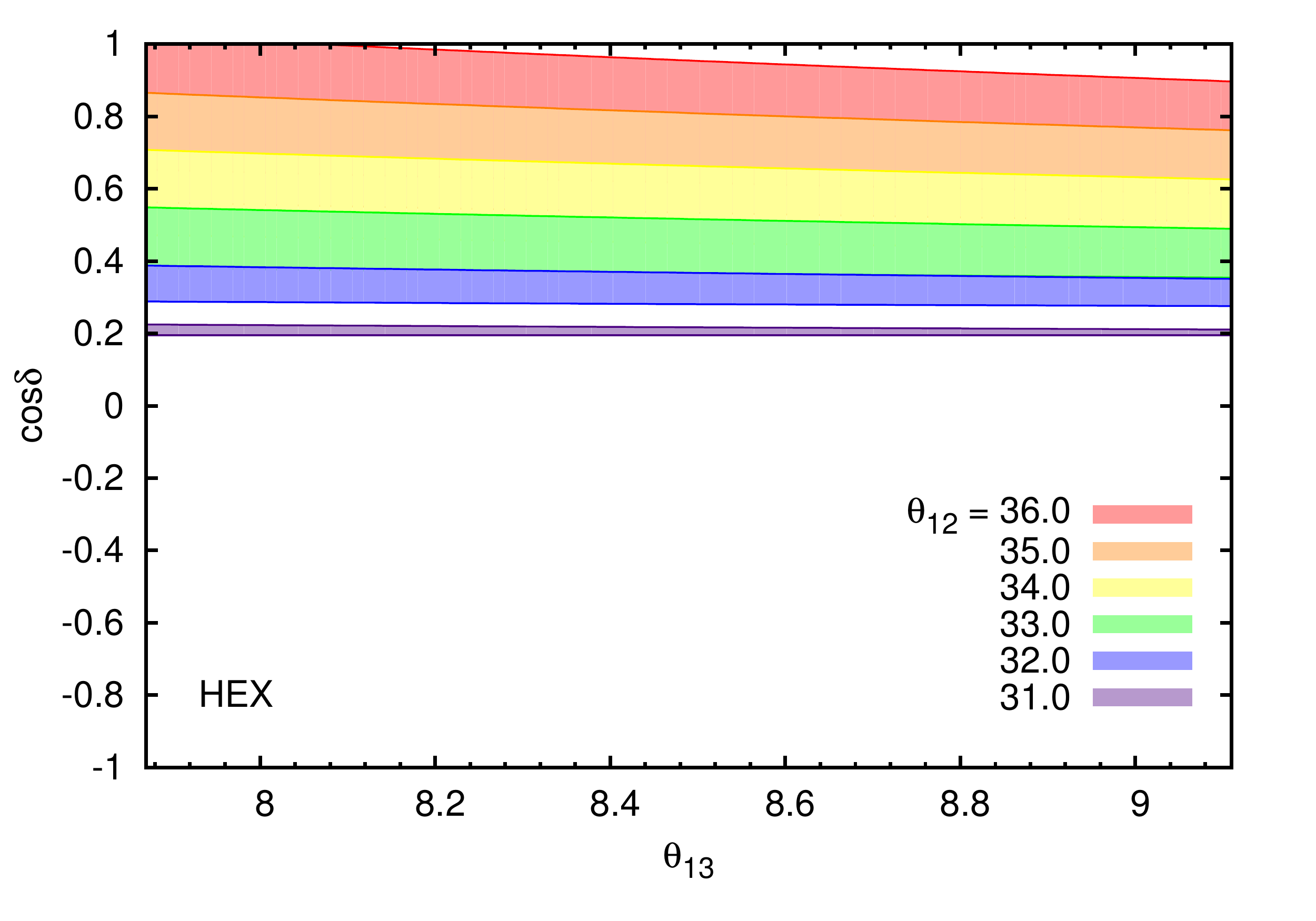}
\end{tabular}
\caption{\label{fig:parameter_space} The predictions for $\cos\delta$ generated
by the solar sum rules for \BM and \TBM (top row), \GR and \GRnew (middle row),
\DGR and \HEX (bottom row). In each plot, the true value of $\theta_{13}$ is
given by the abscissa, the value of $\theta_{12}$ is denoted by the colour of
the band, and the width of the band is generated by varying $\theta_{23}$ over
its $3\sigma$ allowed interval.  }
\end{figure*}
Models which apply $\theta_{12}^e$ and $\theta^e_{23}$ charged-lepton
corrections to a neutrino mixing matrix defined by $\theta_{12}^\nu$ and
$\theta_{23}^\nu$ result in a single constraint given by~\eref{sol4}. For a
given choice of $\theta_{12}^\nu$ this formula may only predict physical values
for $\cos \delta$ in a subregion of the allowed interval $I_{3\sigma}$. In
\figref{fig:parameter_space} we show the regions in which the sum rule makes a
consistent prediction for $\cos\delta$ (coloured bands) for six different
models. In all panels, $\theta_{13}$ is given by the abscissa, $\theta_{12}$ is
denoted by the different coloured bands whose width is generated by varying
$\theta_{23}$ over its range in $I_{3\sigma}$. For bimaximal mixing (\BM) in
the neutrino sector ($\theta_{12}^\nu = 45^\circ$) the only regions of
parameter space for which we find a consistent prediction require a large value
of $\theta_{13}$, a large negative value of $\cos\delta$ and a large value of
$\theta_{12}$. This is easily understood from the linearised relation shown in
\eref{eq:linearSSR}, where the leading-order prediction of $\theta_{12}^\nu =
45^\circ$ must receive large negative corrections to be brought in agreement
with the global data. For tribimaximal mixing (\TBM) in the neutrino sector
($\theta_{12}^\nu = 35.3^\circ$), smaller values of $\theta_{13}$ are allowed,
and all points in $I_{3\sigma}$ lead to consistent predictions. The predicted
values of $\cos\delta$ show only a slight dependence on the true value of
$\theta_{13}$ and $\theta_{23}$, lying between
$-0.7\lesssim\cos\delta\lesssim0.2$.
We consider two models referred to as golden ratio mixing: \GR and \GRnew.  For
\GR, all values of $\theta_{12}$, $\theta_{13}$ and $\theta_{23}$ in
$I_{3\sigma}$ allow for a consistent definition of $\cos\delta$, whereas for
\GRnew we require a small value of $\theta_{12}$, a large value of
$\theta_{13}$ and large positive $\cos\delta$. In \figref{fig:parameter_space},
we see that \GR predicts mostly positive values of $\cos\delta$ with
$0\lesssim\cos\delta\lesssim0.7$. The small region of parameter space in which
\GRnew is consistent with the data is analogous to the allowed regions of the
BM pattern; however, the predictions of $\cos\delta$ for these two models are
distinct. 
The bottom row in {\protect\figref{fig:parameter_space}} shows the possible
predictions for the two patterns related to dihedral symmetries: {\protect\DGR}
and {\protect\HEX}, which are associated with D$_{10}$ and D$_{12}$,
respectively. These models make similar predictions to {\protect\TBM} and
{\protect\GR}, and they give physical values of $\cos \delta$ over the whole
range $I_{3\sigma}$. Our results, shown in
{\protect\figref{fig:parameter_space}}, are in agreement with an independent
survey of charged-lepton corrections presented for the cases of {\protect\TBM},
{\protect\GR}, {\protect\DGR} and {\protect\HEX} in
{\protect\refref{Girardi:2014faa}}.

\begin{figure*}[t]
\centering
\includegraphics[width=0.8\linewidth]{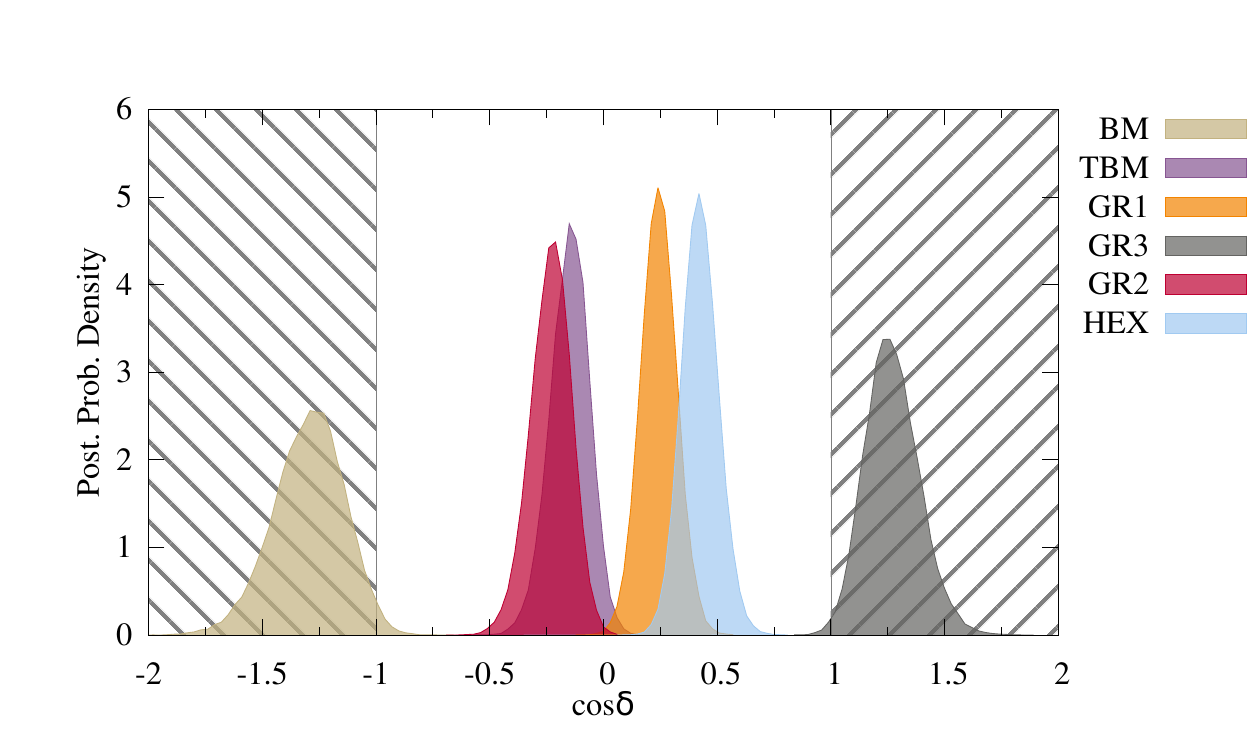}
\caption{\label{fig:MCcosdelta}Posterior probability density functions for
$\cos\delta$ for each of the solar sum rules considered in
\secref{sec:predictions}. The patterned regions are unphysical, which shows
that the \BM and \GRnew sum rules could only be consistent with the known data
if there is a significant deviation from the current best-fit values.}
\end{figure*}

So far we have considered all points in $I_{3\sigma}$ on equal footing.
However, the corners of this parameter space are arguably less likely: they
require large deviations from the current best-fits in multiple parameters. To
fully account for this effect we would need a measure of the degree of
correlation amongst the parameters inferred from global fits. In
{\protect\figref{fig:MCcosdelta}}, in the absence of this information, we show
the posterior probability density functions  for $\cos\delta$
assuming that the likelihood functions for the squared sines of
the mixing angles are given by independent Gaussian distributions centred on
the current best-fit values and with the widths of the global minima. We take a
flat prior in $\sin^2\theta_{ij}$, although we have checked that flat priors in
$\theta_{ij}$ do not significantly change the result. This helps to see the
most reasonable predictions produced by each sum rule if the parameters take
values close to their current best-fits.

In summary, we find that of the four patterns well motivated by symmetry
({\protect\BM}, {\protect\TBM}, {\protect\GR} and
{\protect\GRnew}) only {\protect\TBM} and {\protect\GR} are consistent in a
reasonable part of the parameter space. The predictions associated with
{\protect\BM} and {\protect\GRnew} are only consistent in the far corners of
the $3\sigma$ intervals, where they predict maximal values of $|\cos\delta|$.
For the rest of this work, we shall assume that the solar sum rules derived
from {\protect\BM} and {\protect\GRnew} are excluded.

\begin{figure*}[t]
\begin{tabular}{c c}
\includegraphics[width=0.49\linewidth]{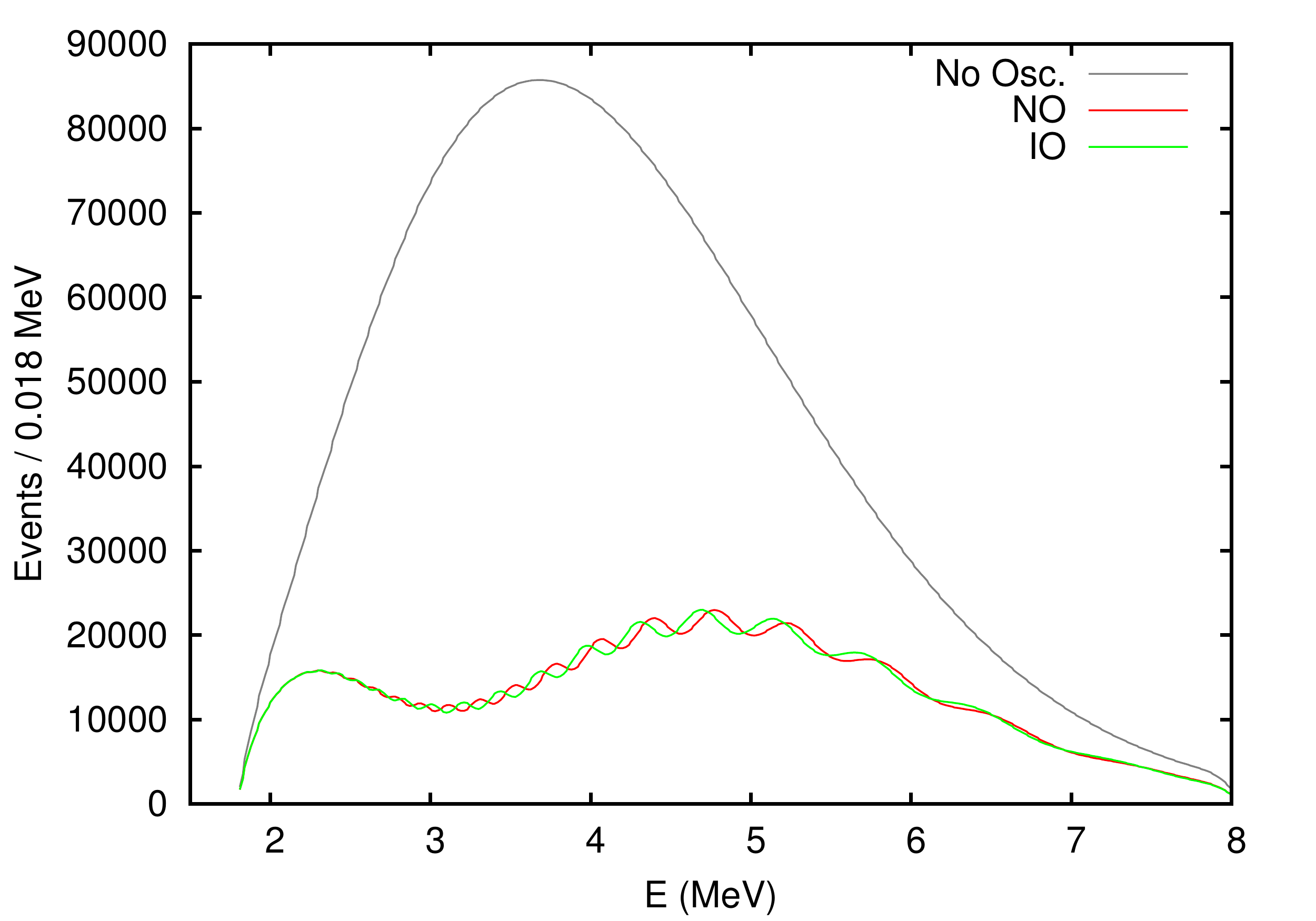} & 
\includegraphics[width=0.49\linewidth]{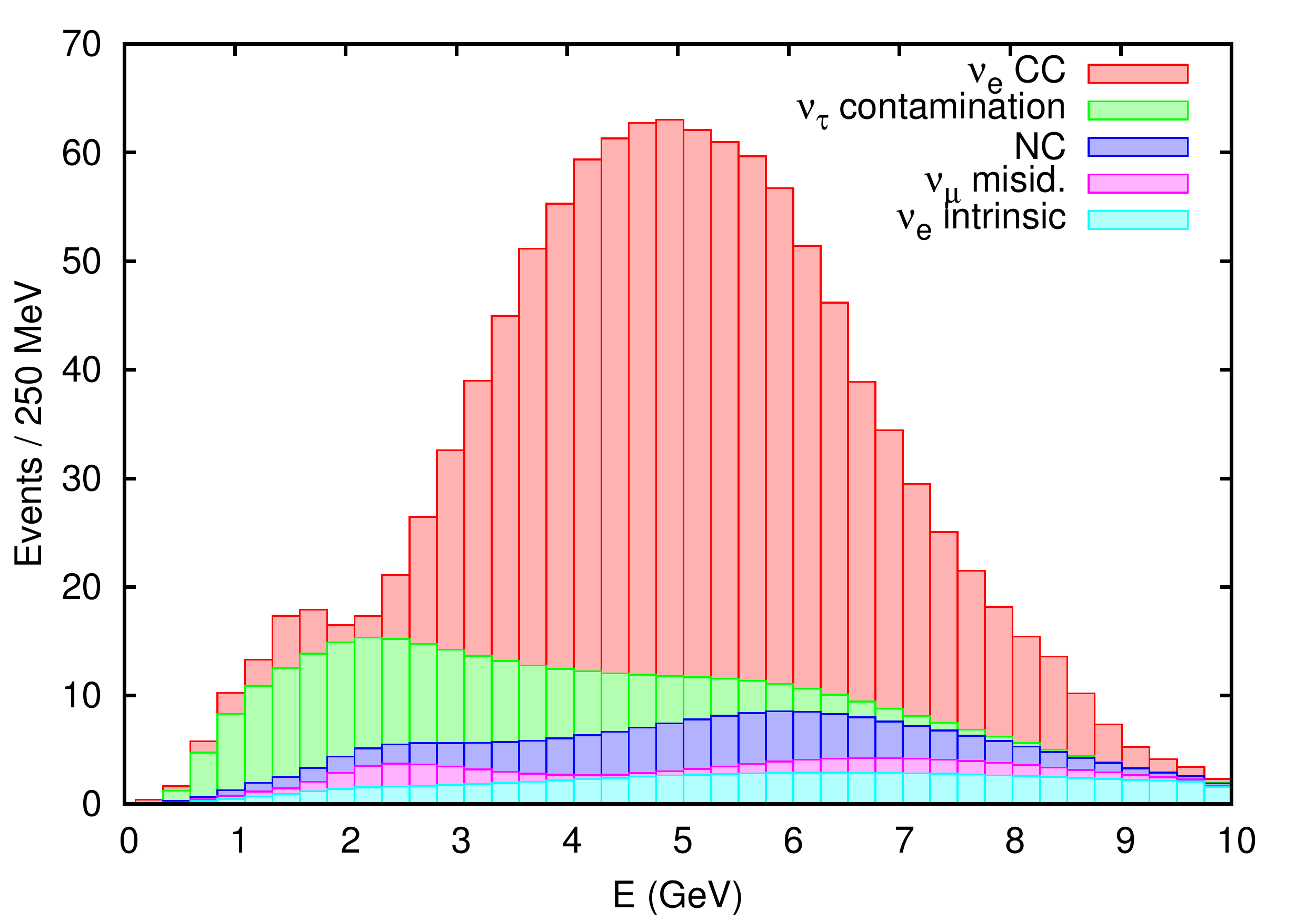} 
\end{tabular}
\caption{\label{fig:LBNO_JUNO_spectra}Example spectra from our simulations of 
the medium-baseline reactor experiment (left) and a wide-band superbeam (right).
}
\end{figure*}

\subsection{\label{sec:simulations}Simulation details}

We simulate the combination of a medium-baseline reactor (MR)
experiment and a wide-band superbeam (WBB). This combination of experiments is
particularly interesting for the investigation of solar sum rules as MR is
expected to improve the current knowledge on $\theta_{12}$,
whilst the superbeam should allow $\delta$ to be constrained at a significant
level for the first time. There are two proposals for a MR with comparable
designs, JUNO and RENO-50, and also two candidates for a next generation WBB,
LBNE and LBNO. Both MRs and WBBs have similar performance targets; however, to
keep our simulations concrete and relevant to experimental work, we will base
our simulations on the JUNO and LBNO designs, and in this subsection we will
discuss the details of our simulations of these facilties. We would like to
stress that this is a purely illustrative choice, and any combination of a MR
and WBB can be expected to perform similarly.

\subsubsection{JUNO}

The Jiangmen Underground Neutrino Observatory (JUNO) is a proposed reactor
neutrino experiment \cite{Wang:2013aa}, whose primary goal is to measure the
neutrino mass hierarchy by observing the subtle shifts that it induces on the
fast subdominant oscillations in the $\overline{\nu}_e$ disappearance
probability \cite{Petcov:2001sy,Qian:2012xh,Ge:2012wj,Li:2013zyd}.  Alongside
the study of the mass hierarchy, this facility has the potential to
significantly improve our measurements of $\theta_{12}$, $\Delta m^2_{21}$ and
$\Delta m^2_{31}$ to a precision of less than $1\%$ \cite{Wang:2013aa}. 

The JUNO experiment derives its flux from twelve nearby reactors, ten of these
are at a distance of around $50$~km from the detector with powers of either $2.9$
or $4.6$~GW, the remaining reactors are much further away at $215$~km and
$295$~km both with powers of $17.4$~GW \cite{Li:2013zyd}. JUNO's detector is
assumed to be a $20$~kton liquid scintillator detector. To measure the fast
oscillations which generate the mass hierarchy sensitivity, JUNO must have very
strong energy resolution capabilities.   A linear energy dependent resolution
of $\Delta E / E = 0.03/\sqrt{E/\mathrm{MeV}}$ is assumed in our simulations
following the design target \cite{Wang:2013aa}.  Non-linearities in the energy
resolution are known to be a possible source of limitations for such an
experiment \cite{Qian:2012xh,Capozzi:2013psa}; however, as these effects are
not as relevant for the precise determination of $\theta_{12}$ we assume that
these effects can be controlled to a negligible level by \emph{in situ}
measurements, and omit them from our simulations. Our spectrum is normalised to
produce $10^5$ total events after $6$~years.

In \refref{Grassi:2014hxa} it has been pointed out that the cosmogenic
muon background for the next generation of large volume reactor neutrino
oscillation experiments with relatively small overburdens is sufficiently large
as to render the KamLAND muon cuts inapplicable whilst preserving a reasonable
active period. This will be a particular problem for the delicate measurement
of the neutrino mass hierarchy, but it will certainly influence the final
sensitivity to $\theta_{12}$ as well. As it stands, it is unclear how the
proposed experiments will circumvent this background. However, for the
successful measurement of the mass hierarchy it must be addressed by some
means, and in the current work, we assume that this background has been brought
under control by a modification of the design or analysis. Even though this
means that our work overestimates the sensitivity to $\theta_{12}$, it should
not significantly alter our analysis of solar sum rules, which for the most
part, only require a precision of $\theta_{12}$ at the percent level to be
effective. 

An illustrative event spectrum is shown in \figref{fig:LBNO_JUNO_spectra} for
JUNO in both mass hierarchies (with and without oscillations). Our simulation
agrees with the predicted performance in \refref{Wang:2013aa}; in particular,
our simulation provides an independent precision on $\theta_{12}$ of around
$0.6$\%.

\subsubsection{Wide-band superbeam}

A superbeam is the extrapolation of conventional neutrino beam production
methods to more intense beams and larger detectors. The source neutrino
beam is produced at an accelerator, which collides protons with a fixed target
generating a spray of mesons, predominately $\pi^\pm$. Magnetic focusing
selects mesons of a given charge and the decay of these particles produces a
beam of neutrinos. The flavour profile is mostly $\nu_\mu$ (for focused
$\pi^+$) and $\overline{\nu}_\mu$ (for focused $\pi^-$); although, there is a
small contamination from subdominant meson decay modes which leads to an
intrinsic background of $\nu_e$ and $\overline{\nu}_e$ at the sub-percent level.

Our model of the wide-band superbeam is based upon the Long Baseline Neutrino
Experiment (LBNE) {\protect\cite{Adams:2013qkq}} and Long Baseline Neutrino
Oscillations (LBNO) \protect{\cite{Agarwalla:2013kaa}} proposals for on-axis
superbeams with baselines of around $1000$--$2000$~km.  The on-axis
orientation ensures that the beam has a wide spectrum and allows the
oscillation probability to be tested over a range of values of $L/E$,
mitigating degeneracies and improving precision.
Both of these experiments aim to determine the mass hierarchy and the
CPV phase $\delta$ through the precise measurement of the appearance channel 
probabilities $P(\nu_\mu\to\nu_e)$ and $P(\overline{\nu}_\mu\to\overline{\nu}_e)$.  
LBNE and LBNO have a comparable physics reach, which is ultimately dependent
on the precise programme of upgrades available, and we base our simulations on
LBNO as described in \refref{Agarwalla:2013kaa}. The LBNO design features as
its first phase a $700$~kW beam, a baseline distance of $2300$~km between CERN
and the Pyh\"{a}salmi mine in Finland, and a $20$~kton detector based on
liquid argon time-projection chamber
technology \cite{Agarwalla:2013kaa,Stahl:2012exa}. Our simulation of this
facility uses the fluxes provided by \refref{Longhin:2012ae}, and propagates
the neutrinos through a constant density background of $3.2$~g/cm$^3$. We
consider both the appearance $\nu_\mu\to\nu_e$
($\overline{\nu}_\mu\to\overline{\nu}_e$) and muon disappearance channels
$\nu_\mu\to\nu_\mu$ ($\overline{\nu}_\mu\to\overline{\nu}_\mu$).  The
background to the appearance channel is given by the intrinsic $\nu_e$
component of the beam, misidentified $\nu_\mu$ events at a rate of $1\%$,
$2\%$ of neutral current events and events arising from $\tau$-contamination:
the production of $\tau^\pm$ leptons in the detector which quickly decay to
$e^\pm$. These $\tau$ events have been implemented via a custom migration
matrix which maps the spectrum of incoming $\nu_\mu$ $(\overline{\nu}_\mu)$
onto the resultant $e^-$ ($e^+$) post-decay distribution. 
We have normalised our number of events to match the tables simulated in
\refref{Agarwalla:2013kaa}, which assumes a total of $10^{21}$ protons on
target corresponding roughly to $10$ years of run time, but we consider masses
of $35$~kton and $70$~kton to account for a reasonable range of possible
detectors, according to the LBNE and LBNO phased designs and upgrade
programmes \cite{Agarwalla:2013kaa, Adams:2013qkq}. An
illustrative spectrum decomposed into its background components can be seen in
the right panel of \figref{fig:LBNO_JUNO_spectra}. We see a close agreement of
form for most of our backgrounds when our spectrum is compared with the spectra in
\refref{Agarwalla:2013kaa}. The only notable deviation is in the shape of our
neutral current background, but this small difference is not expected to effect
the general conclusions of the current work.

\subsection{Simulation results}

\begin{figure}[tb]
\centering
\includegraphics[angle=270,width=0.5\linewidth,clip,trim=14 2 22 33]{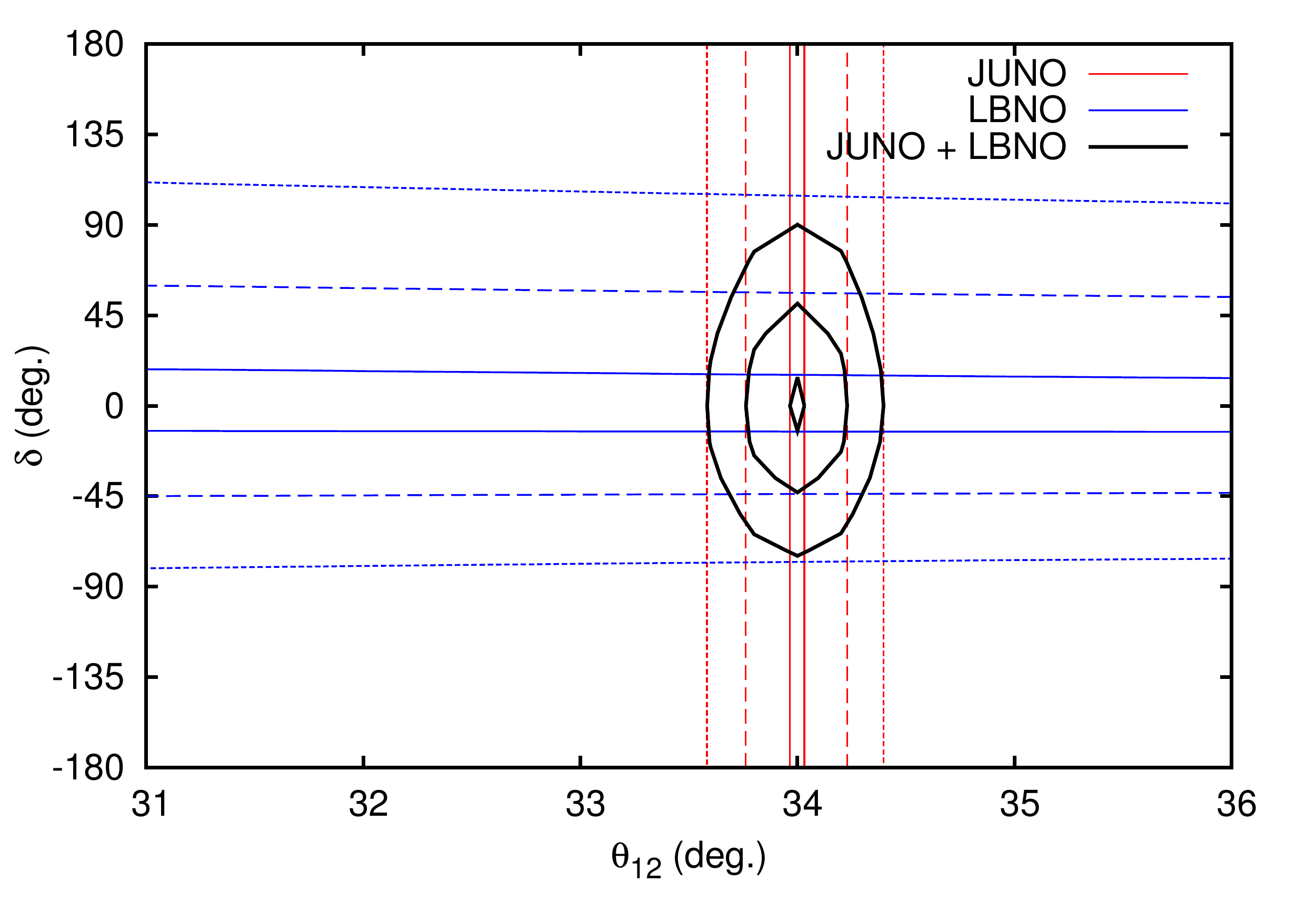}
\caption{\label{fig:JUNO_LBNO}The independent and combined constraints on the
parameters $\delta$ and $\theta_{12}$ for an illustrative point in parameter
space for the JUNO and LBNO experiments. In each experimental arrangement, the
three lines correspond to $1$, $3$ and $5\sigma$ significance.}
\end{figure}

Our simulation combines the expected data from a long-baseline wide-band
superbeam (WBB) experiment, modelled after LBNO although it also provides a
good estimate of the performance of LBNE, and a medium-baseline reactor (MR)
experiment with a baseline around $60$~km, modelled on JUNO.
As discussed previously, these facilities are expected to  provide
complementary constraints on the parameters relevant for the solar sum rule.
This synergy can be seen in \figref{fig:JUNO_LBNO}, where we show the
independent constraints on $\delta$ and $\theta_{12}$ provided by the LBNO and
JUNO experiments and their combination. As expected, the precision is dominated
by JUNO for $\theta_{12}$ and by LBNO for $\delta$.

To determine the allowed regions for a given sum rule we generate the expected
event rates for each set of true parameters of interest. We then compute the
set of hypothetical parameters which maximise the likelihood function under two
different hypotheses: first, assuming no constraints on the parameter space,
which finds the best-fitting point, and secondly, whilst imposing the
constraint of the sum rule on our hypothesised parameters. When the constrained
best-fit has a likelihood significantly below the unconstrained best-fit, we
conclude that the sum rule can be excluded.
During the maximisation process, we include priors on the oscillation
parameters to account for the external constraints of the global data. If not
mentioned explicitly, we assume the following values for the true parameters
\[ \theta_{12} = 33.48^\circ, \quad \theta_{13} = 8.50^\circ
\quad\text{and}\quad \theta_{23} = 42.3^\circ,\]
and the mass splittings 
\[  \Delta m^2_{21} = 7.50\times10^{-5}~\text{eV}^2 \quad\text{and}\quad \Delta
m^2_{31} = 2.46\times10^{-3}~\text{eV}^2. \]
For the prior constraints, we assume a $2\%$ ($2.4\%$) uncertainty on $\Delta
m^2_{31}$ ($\Delta m^2_{21}$) and uncertainties of $2.3\%$, $2.4\%$ and $6\%$ on
$\theta_{12}$, $\theta_{13}$ and $\theta_{23}$, respectively. These are chosen
to be in agreement with recent best-fits \cite{Gonzalez-Garcia:2014bfa}.  We
assume normal hierarchy in all of our simulations, although it has been checked
that the results are not particularly sensitive to this choice. Our
simulations are implemented using the GLoBES package \cite{Huber:2004ka,
Huber:2007ji}. 

\begin{figure}[t]
\begin{tabular}{c c}
\includegraphics[width=0.49\linewidth]{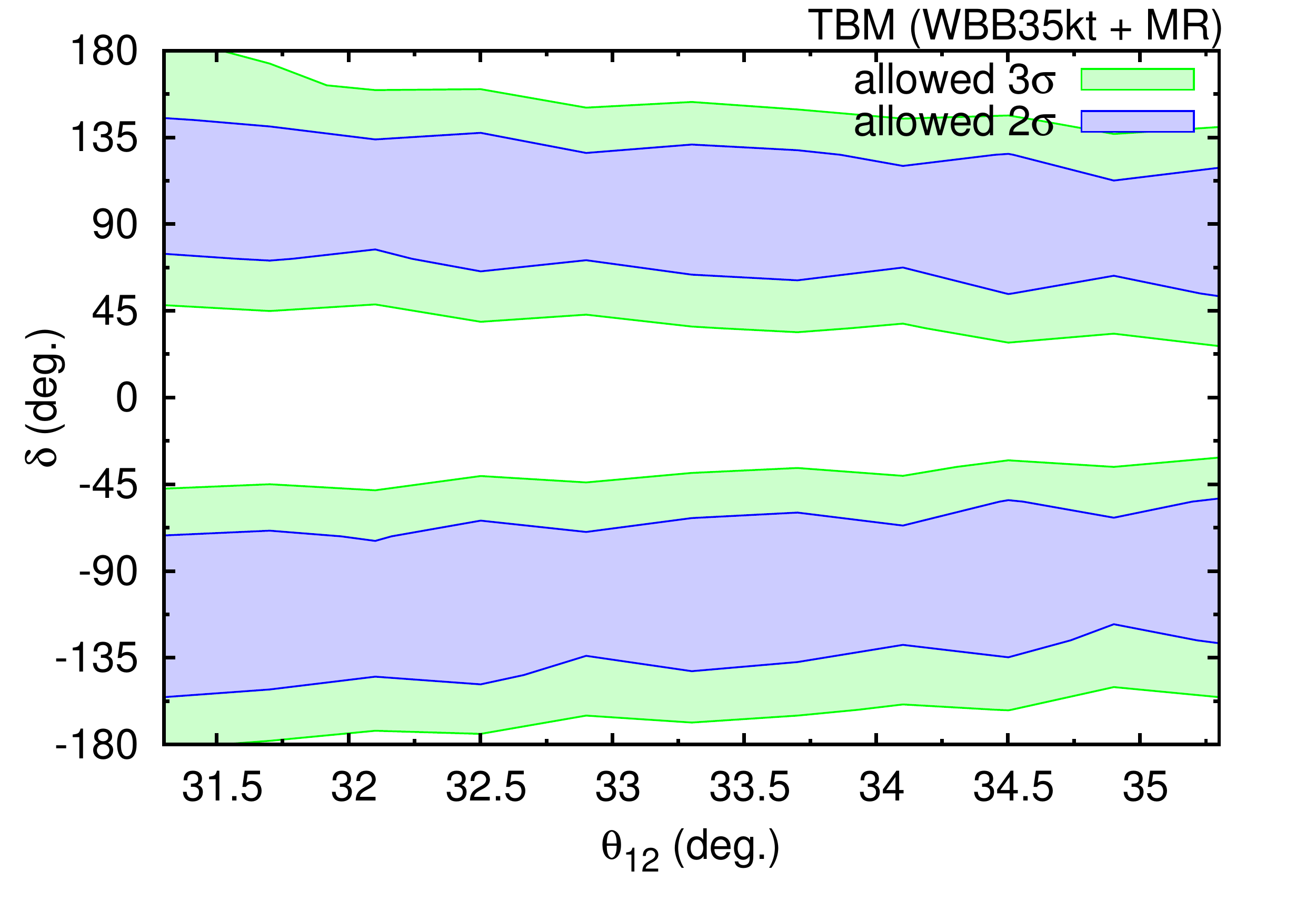} &
\includegraphics[width=0.49\linewidth]{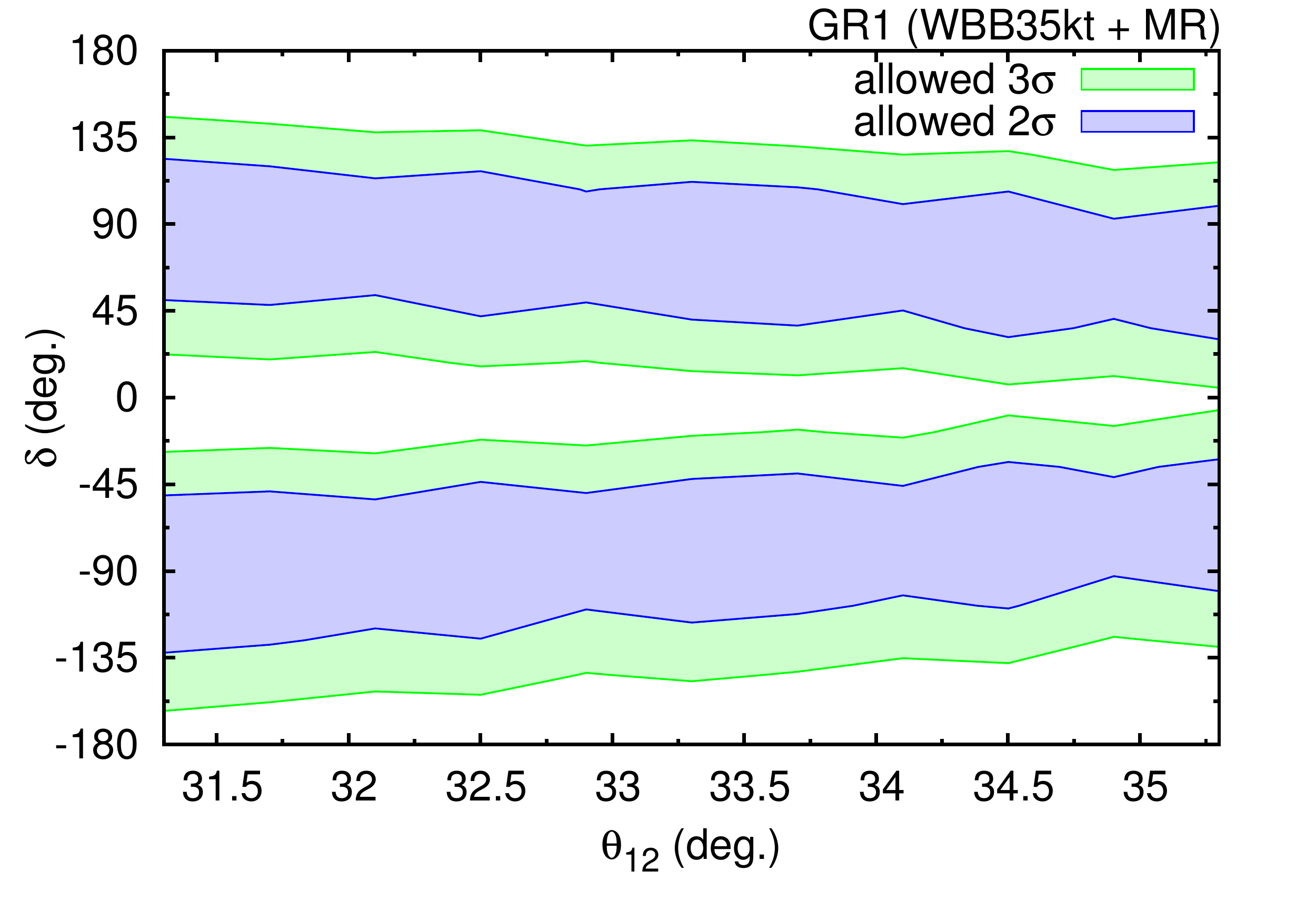} \\
\includegraphics[width=0.49\linewidth]{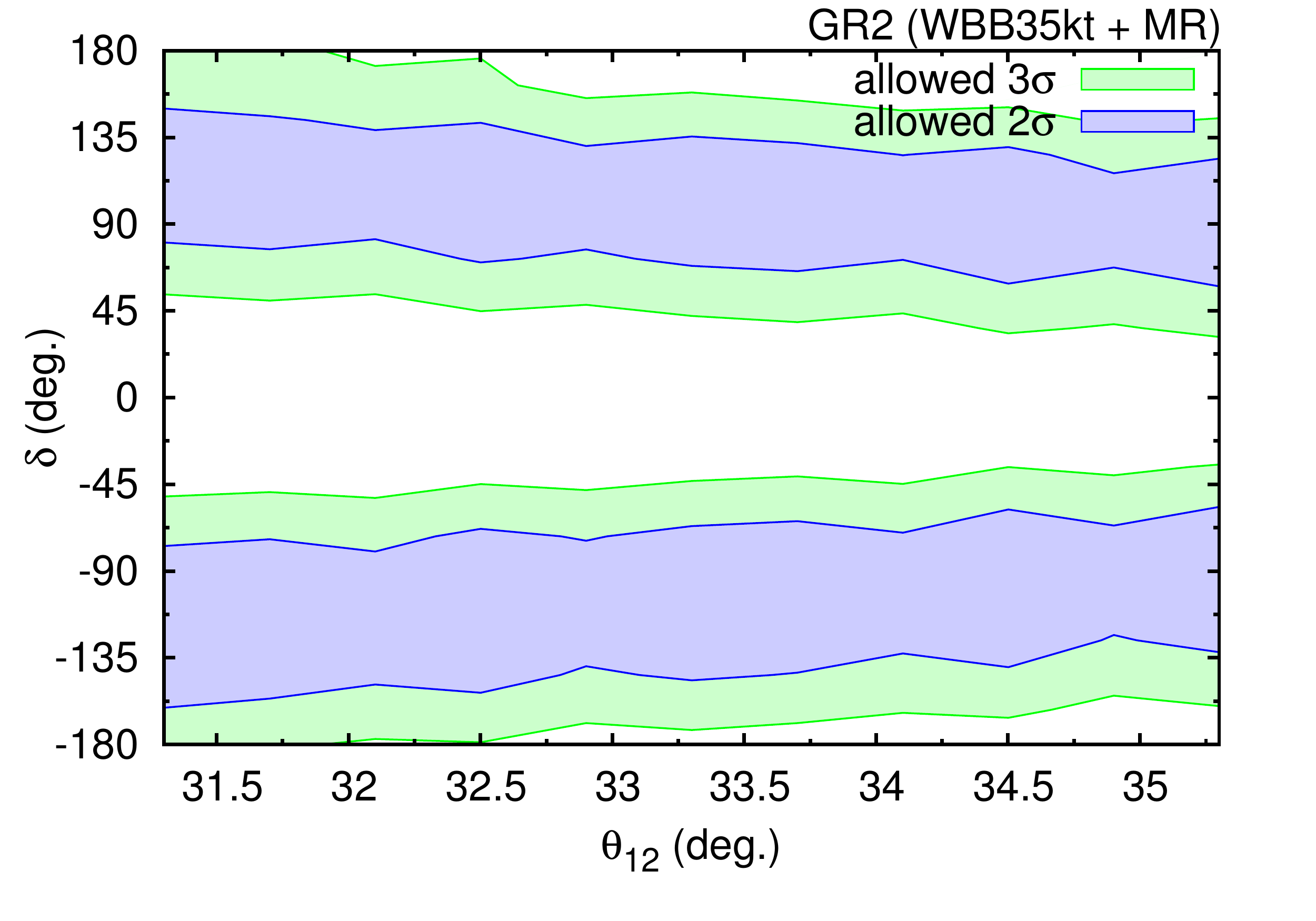} &
\includegraphics[width=0.49\linewidth]{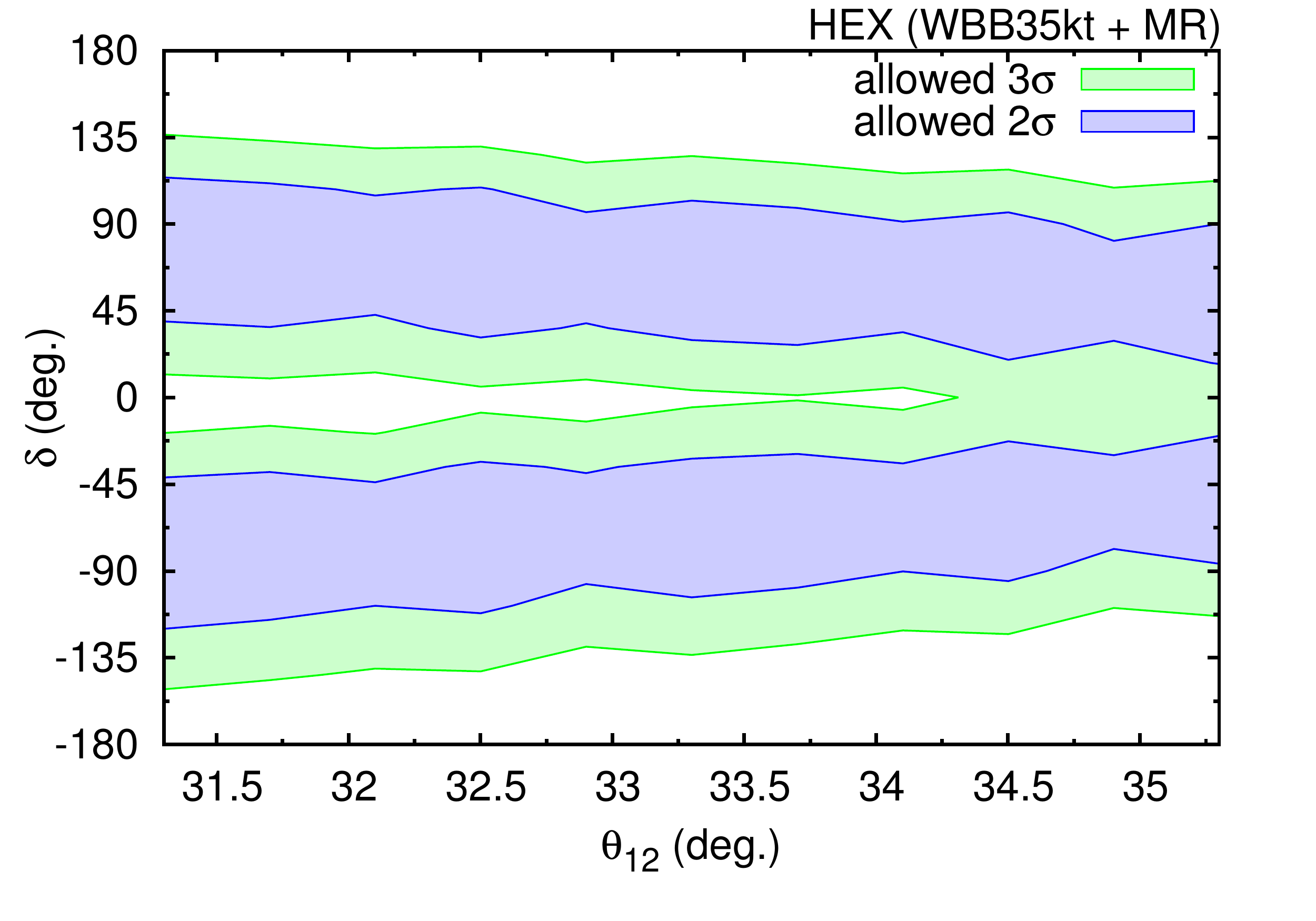} 
\end{tabular}
\caption{\label{fig:solar_sum_rules}The
allowed regions of true parameter space in the $\theta_{12}-\delta$ plane
for \TBM (left, top row), \GR (right, top row), \DGR (left, bottom row) and \HEX (right, bottom row) after $6$ years of data taken by a medium-baseline reactor experiment (MR) and $10$ years by a wide-band superbeam with $35$~kton detector (WBB35kt).}
\end{figure}

In \figref{fig:solar_sum_rules}, we show the results of our simulations of MR
and WBB for the measurement of the solar sum rules defined by
{\protect\eref{sol4}}.  The four models shown are the tribimaximal (TBM) model
with $\theta^\nu_{12}=\arcsin\frac{1}{\sqrt{3}}$, the golden ratio model with
$\theta^\nu_{12}=\arctan\frac{1}{\varphi}$, and the two dihedral models,
{\protect\DGR} and {\protect\HEX}, with $\theta^\nu_{12}=\frac{\pi}{5}$ and
$\theta^\nu_{12}=\frac{\pi}{6}$, respectively. In the left panel on the top
row, we see the regions of true parameter space for which the TBM model cannot
be excluded at $2$ and $3\sigma$ significance. 
As we saw in {\protect\figref{fig:parameter_space}}, all four of the
models under consideration here make consistent predictions for all values of
$\theta_{12}$. The plots predict a similar fraction of true parameter space in
which these models can be excluded: at $2\sigma$ the models can be excluded in
around $65\%$ of the parameter space, while at $3\sigma$ this drops to around
$26\%$. These plots show only a mild dependence on the true value of
$\theta_{12}$. Comparing the panels on the top row, which show the two models
well motivated by symmetry, we see that the smaller value of $\theta^\nu_{12}$
in the {\protect\GR} model compared to the {\protect\TBM} prediction generates
a smaller prediction for $|\delta|$. Although, there is significant overlap
between the allowed regions of the two models, which means it is unlikely that
they could be distinguished with the $35$~kton detector.
The bottom row shows the two dihedral predictions ({\protect\DGR} on the left,
and {\protect\HEX} on the right). For {\protect\HEX}, the smaller value of
$|\delta|$ leads to the $3\sigma$ regions merging at $\delta\approx0$. In
practice, this means that for larger values of $\theta_{12}$, it is easier to
exclude the model: for $\theta_{12}=35.3^\circ$, we see the fraction of
parameter space for which the model can be excluded increase to around $38\%$
at $3\sigma$.

Although distinguishing between these models will be challenging, each can be
excluded for a reasonable region of the parameter space through the combination
of data from a MR and WBB experiment. In general, we can also point out that an
observation of an extreme value of $\cos\delta$ would allow all of these models
to be excluded: a true value of $|\cos\delta|=1$ or $\cos\delta=0$ would 
disfavour all models at $2\sigma$ and for most of the parameter space exclude
them at $3\sigma$.

\begin{figure}[t]
\begin{tabular}{c c}
\includegraphics[width=0.49\linewidth]{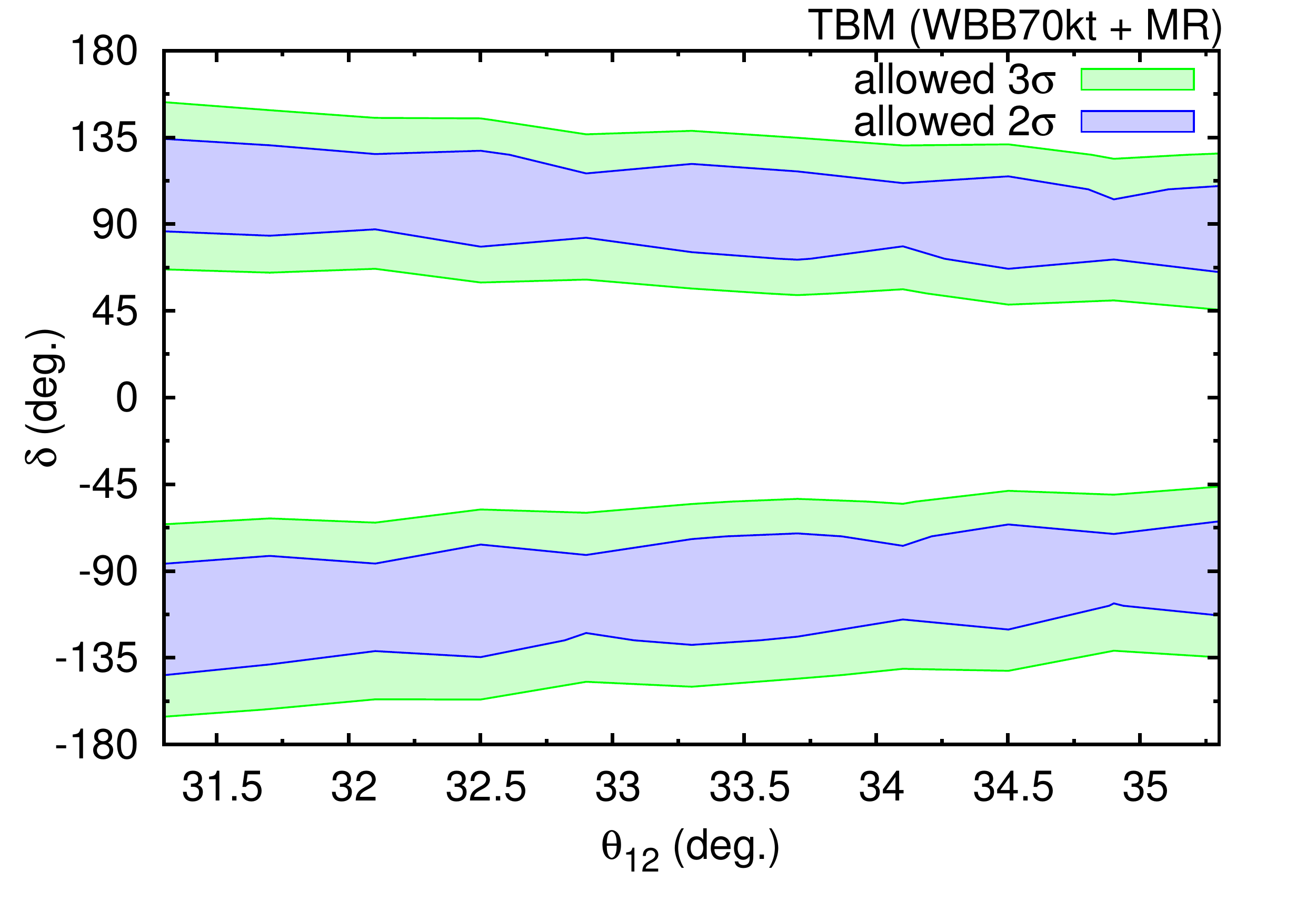} &
\includegraphics[width=0.49\linewidth]{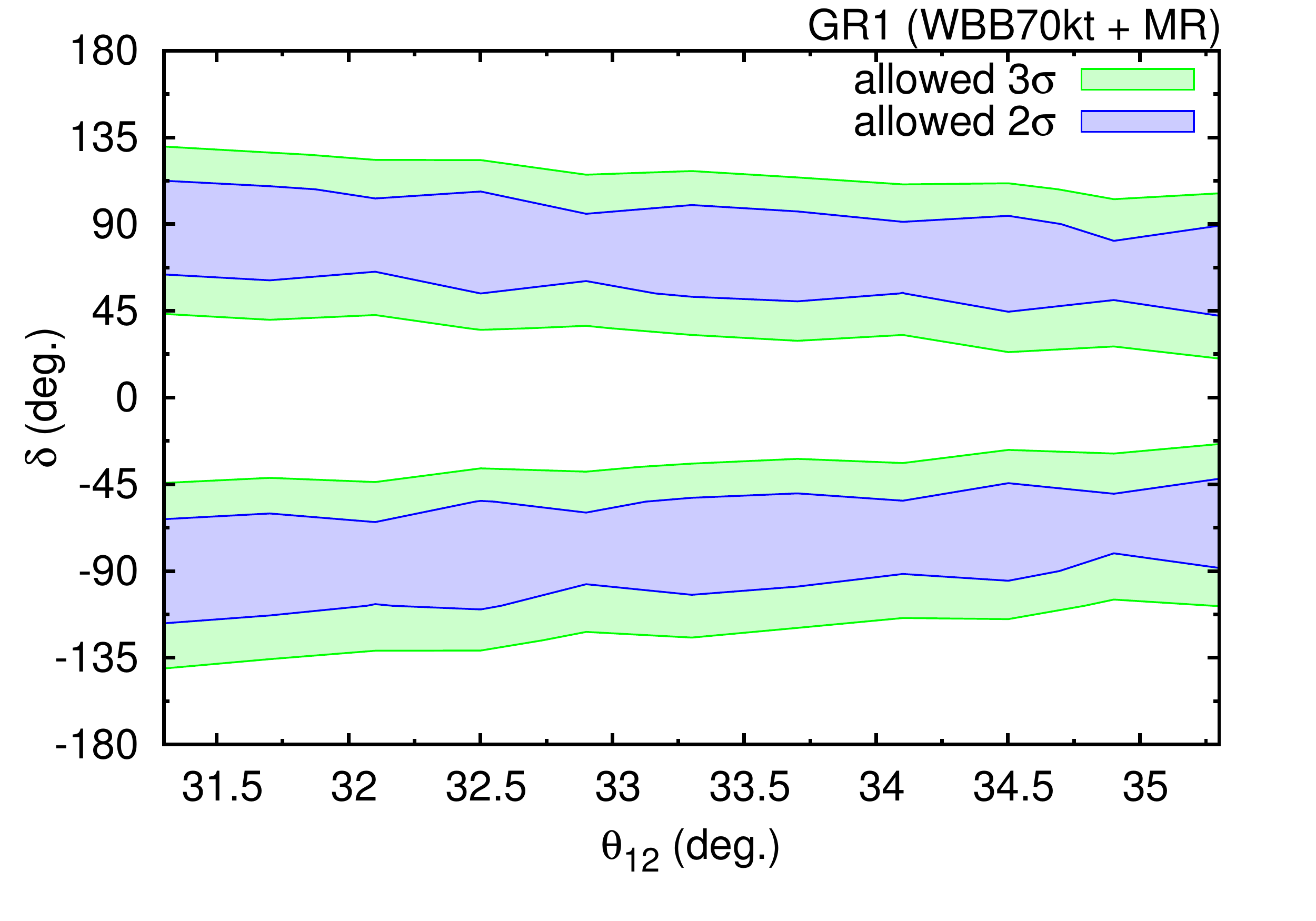} \\
\end{tabular}
\caption{\label{fig:solar_sum_rules_70}The 
allowed regions of true parameter space in the $\theta_{12}-\delta$ plane
for \TBM (left) and \GR (right) after $6$ years of data taken by a medium-baseline reactor experiment (MR) and $10$ years by a wide-band superbeam with an upgraded $70$~kton detector (WBB70kt).}
\end{figure}

Upgrading the WBB experiment would allow the discovery potential
to be significantly extended. In {\protect\figref{fig:solar_sum_rules_70}}, we
show the effect of increasing the detector mass to $70$~kton. This doubling of
statistics allows the precision on $\delta$ to increase, which leads to larger
exclusion areas. We show the results for {\protect\TBM} and {\protect\GR},
where the exclusion regions are now around $71\%$ of the true parameter space
at $2\sigma$ and $48\%$ at $3\sigma$. This corresponds to an $80\%$ increase
in the $3\sigma$ exclusion region. Clearly, to fully understand the potential for this
measurement, the foreseen programme of upgrades will play an important role.

\section{\label{sec:beyond}Beyond $\theta^{\nu}_{13}=0$}
Neutrino mixing patterns in which $\theta^{\nu}_{13}\neq 0$ have also been
predicted in the literature, and can likewise give rise to solar sum rules.
However, \eref{sol4} does not apply in such scenarios, and we must work on a 
case by case basis. Examples of patterns of this type can be found in the
fully-specified patterns in \refref{Fonseca:2014koa}. Of these patterns, we
identify 10 unique mixing matrices from the sporadic patterns which meet our
criteria on the mixing angles and have $\theta_{13}^\nu\neq0$, along with an
infinite subset of the family $\mathcal{C}_2$. 

As an example, we shall derive another sum rule from one member of
$\mathcal{C}_2$ characterised by a purely imaginary parameter
$\sigma={i}$, cf. Eq.~(150) of \refref{Fonseca:2014koa}. This pattern
had been previously studied in a grand-unified model of flavour based on the
group $\Delta(96)\times SU(5)$ \cite{King:2012in}, where it was known as the
bi-trimaximal mixing (BTM) pattern,  
\be U^\nu_{\text{BTM}} ~=~ \begin{pmatrix} a_+ & \frac{1}{\sqrt{3}} & a_- \\
-\frac{1}{\sqrt{3}} & \frac{1}{\sqrt{3}} & \frac{1}{\sqrt{3}} \\ a_- &
-\frac{1}{\sqrt{3}} & a_+ \end{pmatrix}, \ee 
with $a_{\pm}=(1\pm \frac{1}{\sqrt{3}})/2$. Multiplication of the charged
lepton mixing ${U_{12}^e}^\dagger$ from the left yields the PMNS matrix (up to
Majorana and unphysical phases) 
\be U ~=~ \begin{pmatrix} a_+c^e_{12} + \frac{1}{\sqrt{3}}
s^e_{12}e^{-i\delta^e_{12}} & \frac{1}{\sqrt{3}}c^e_{12} - \frac{1}{\sqrt{3}}
s^e_{12}e^{-i\delta^e_{12}}& a_-c^e_{12} - \frac{1}{\sqrt{3}}
s^e_{12}e^{-i\delta^e_{12}}\\ a_+s^e_{12}e^{i\delta^e_{12}}
- \frac{1}{\sqrt{3}} c^e_{12}& \frac{1}{\sqrt{3}}s^e_{12}e^{i\delta^e_{12}} +
  \frac{1}{\sqrt{3}} c^e_{12} & a_-s^e_{12}e^{i\delta^e_{12}} +
\frac{1}{\sqrt{3}} c^e_{12}\\ a_-&-  \frac{1}{\sqrt{3}} & a_+ \end{pmatrix}.
\ee 
The two free continuous parameters $\theta_{12}^e$ and $\delta_{12}^e$
will control the four physical parameters $\theta_{12}$, $\theta_{13}$,
$\theta_{23}$ and $\delta$. Therefore we expect two sum rules. They can be
derived easily by comparing the (square of the) absolute value of the $(i,j)$
entry with the corresponding entry in the PDG parameterisation of $U$. From
$U_{\tau3}$ we find the first exact sum rule involving the atmospheric angle 
\be c_{13} c_{23} ~=~ a_+ \ .  \label{eq:bt1} \ee 
Similarly we get from $U_{\tau2}$ 
\be c_{23}^2 s_{12}^2 s_{13}^2 + s_{23}^2 c_{12}^2+ 2 s_{23} c_{23}s_{12}c_{12}
s_{13} \cos \delta ~=~ \frac{1}{3} \ .  \label{eq:bt2a} \ee
Solving \eref{eq:bt1} for $\theta_{23}$ and inserting the result into
\eref{eq:bt2a} gives rise to the sum rule involving $\theta_{12}$,
$\theta_{13}$ and  $\delta$,  
\be \cos \delta ~=~ \frac{c_{13}^2-3a_+^2 s_{12}^2 s_{13}^2 -3c_{12}^2
(c_{13}^2-a_+^2)  }{ 6 a_+  s_{12}c_{12} s_{13} \sqrt{c_{13}^2-a_+^2} } \ .
\label{eq:bt2} \ee

Satisfying both of these constraints simultaneously is very difficult with the
known global data on the mixing angles: $\cos\delta$ is only well defined for
large values of $\theta_{13}$ and small values of $\theta_{12}$, but the
constraint of \eref{eq:bt1} requires that $\theta_{13}$ and $\theta_{23}$
are at the low-valued extremes of their allowed parameter space. 
This tension may be alleviated by introducing further corrections to these
predictions, for example the renormalisation effects.

\section{\label{sec:rge}Renormalisation group corrections}

In this analysis we have ignored the effects of renormalisation group (RG)
corrections to mixing angles.  Although this is generally a good approximation,
it is useful to be aware of the typical magnitudes of such corrections and when
they might be important. In this section, we briefly review such issues.
For previous discussion of RG corrections in this context see \eg~
\protect\refref{Schmidt:2006rb} for a discussion in case of Cabibbo-like
charged lepton correction to BM mixing and charged lepton corrections to TBM in
\protect\refref{Boudjemaa:2008jf}.

In the effective theory the RG correction to $\theta_{12}$,
which is generically the largest, is described by the following
renormalisation group equation~\cite{Antusch:2003kp}
\begin{equation}\label{eq:RGt12Approx}
\dot\theta_{12} \equiv 
\frac{d \theta_{12}}{d\ln (\mu/\mu_0)}
=-\frac{C y_\tau^2}{32\pi^2}\sin2\theta_{12}s_{23}^2
\frac{\left|m_1 e^{i\varphi_1}+m_2
e^{i\varphi_2}\right|^2}{m_2^2-m_1^2}+\mathcal{O}(\theta_{13})\ ,
\end{equation}
with $C=1$ in the Minimal Supersymmetric Standard Model (MSSM) 
and $C=-3/2$ in the Standard Model (SM). 
In the MSSM large values of $\tan \beta = v_u/v_d$ 
lead to an enhancement of the RG running via the Yukawa coupling 
$y_\tau^2=m_\tau^2 (1+\tan^2\beta) /v^2$, where $v=246\,$GeV. 
There is no such enhancement in the SM. 
In order to estimate the size of the RG corrections, we use the exact analytic RG equation for $\theta_{12}$ fixing all parameters at their respective best fit values. We do not include the running of the other parameters.
The running of the solar mixing angle from the electroweak scale to the GUT scale
$\Lambda_\mathrm{GUT}=2\cdot 10^{16}$\,GeV is given to leading order by 
$\Delta \theta_{12}=\theta_{12}(\Lambda_\mathrm{GUT})-\theta_{12}(m_Z) \simeq
\dot\theta_{12}(m_Z) \ln \left(2 \cdot 10^{14}\right)$. It depends on the
absolute neutrino mass scale. 

In the SM the corrections are generally small. Using the exact\footnote{
It turns out that the term proportional to $\theta_{13}$ becomes
relevant in \protect\eref{eq:RGt12Approx} and the exact expression of $\dot \theta_{12}$ has to be used.} one-loop formula for $\theta_{12}$ given in \protect\refref{Antusch:2003kp}, we obtain a conservative estimate by taking a quasi-degenerate mass
spectrum with vanishing phases: 
$\Delta\theta_{12}\approx 0.17^\circ$ ($0.04^\circ$) 
for $m_1=0.1$ eV ($m_1=\sqrt{\Delta m_{32}^2}$).
Experimentally, the expected sensitivity for determining $\theta_{12}$ is
about $0.6\%$~\cite{Wang:2013aa}, i.e. $\Delta \theta_{12} \approx
0.2^\circ$.  In the context of the SM it is therefore justified to neglect RG
effects on the solar sum rules discussed in this paper. 

On the other hand, RG corrections are typically bigger in the MSSM.  We have
considered 4 different cases: normal and inverted mass ordering and a Majorana
phase difference $\Delta \varphi=\varphi_2-\varphi_1$ of $0$ and $\pi$.  With
$\Delta \varphi = 0$, RG effects become more and more relevant for increasing
absolute neutrino mass scales and increasing values of $\tan \beta$. For an
inverted mass ordering, the lower bound on $m_1$ entails relevant RG
corrections for $\tan \beta \gtrsim 2$. In the case of a normal mass ordering,
RG corrections have to be taken into account if $\tan \beta \gtrsim 45$, where
this bound is decreasing with increasing $m_1$, \eg with $m_1 \approx 
\sqrt{\Delta m_{21}^2}$ one has $\tan\beta \gtrsim 10$.
A non-vanishing Majorana phase difference generically suppresses RG running.
For $\Delta \varphi=\pi$, the leading term in \eref{eq:RGt12Approx} is
proportional to $\Delta m_{21}^2/(m_1+m_2)^2$, such that it {\it decreases}
with increasing absolute neutrino mass scale. 
RG corrections to the solar sum rules can be neglected for $\tan \beta \lesssim 35$ provided the neutrino mass spectrum is not quasi-degenerate.

\section{\label{sec:conclusion}Conclusion}

We have presented a succinct derivation of solar lepton mixing
sum rules, arising from simple patterns of neutrino mixing with
$\theta_{13}^\nu=0$, enforced by discrete family symmetry and corrected by a
rather generic charged lepton mixing matrix, assuming only that
$\theta_{13}^e=0$. From our derivation we have expressed the result as the
ratio of the absolute magnitude of two PMNS matrix elements, given in terms of
$\theta_{12}^{\nu}$, namely $| U_{\tau 1}|/ | U_{\tau 2} | =t^{\nu}_{12}$.
When expanded in terms of the three PMNS mixing angles and the CPV oscillation
phase $\delta$, the resulting solar sum rule may be cast in terms of a 
prediction for $\cos \delta$ which depends only on $\theta_{12}^{\nu}$.

We have considered in detail the resulting solar mixing sum rules, arising from
four particularly well-motivated cases of neutrino mixing, which
can be derived from discrete family symmetries, namely:
{\protect\BM} mixing where $s^{\nu}_{12}=1/\sqrt{2}$,
{\protect\TBM} mixing where $s^{\nu}_{12}=1/\sqrt{3}$, and two patterns based
on versions of golden ratio mixing including \GR with
$t^{\nu}_{12}=1/\varphi$ and \GRnew with $c^{\nu}_{12}=\varphi /\sqrt{3}$, where
$\varphi=(1+\sqrt{5})/2$ is the golden ratio.  We have also fully discussed two
leading-order patterns which have been invoked in the literature, but which
cannot be enforced by any simple discrete family symmetry, called \DGR with
$\theta^{\nu}_{12}=\pi /5$ and \HEX mixing with $\theta^{\nu}_{12}=\pi /6$. 

It turns out that two of the above six cases, namely \BM and \GRnew, are almost
excluded by current data, so in the phenomenological study, we
have focused on the remaining four viable cases, namely \TBM, \GR,
\DGR and \HEX, of which only the first two (\TBM and \GR) are well founded by
symmetry arguments.  The predictions for $\cos \delta$ for all these cases are
summarised in {\protect\figref{fig:MCcosdelta}}.  For the four viable cases, we
performed a simulation of a next-generation superbeam experiment, based on
LBNO, and a future reactor experiment, based on JUNO, to see how well the sum
rules can be tested. For example, in \figref{fig:solar_sum_rules} we show the
allowed regions of true parameter space in the $\theta_{12}-\delta$ plane,
following a 6-year medium-baseline reactor experiment and a
decade of running with a WBB and a 35 kton detector.

We have seen that the ability to constrain solar sum rules relies crucially on
the complementary sensitivities of both reactor and superbeam experiments, and
that, acting together, these facilities will be capable of significantly
restricting the allowed parameter space of the models associated with solar
mixing sum rules.  It is possible that such experiments could exclude
all of the models considered in this paper, which would be the
case if, for example $\delta \approx 0$. If this occurs, the
theoretical approach of explaining lepton mixing as a result of charged-lepton
corrections to simple symmetry-driven patterns of neutrino mixing
would become strongly disfavoured.

\section*{Acknowledgements}
We acknowledge support from the European Union FP7 ITN-INVISIBLES (Marie Curie
Actions, PITN- GA-2011-289442). 
The work of CL is supported by the Deutsche Forschungsgemeinschaft within the
Research Unit FOR 1873 (Quark Flavour Physics and Effective Field Theories).
MS acknowledges support by the Australian Research Council.
This work has been additionally supported by the European Research Council
under ERC Grant ``NuMass'' (FP7-IDEAS-ERC ERC-CG 617143).

%%%%%%%%%%%%%%%%%%%%%%%%%%%%%%
\appendix \section{\label{sec:simpApp}Simple approximation to the mixing sum
rule}
In this appendix we show that the exact sum rule in \eref{sol4} reduces to the
well-known leading-order sum rule to an accuracy of a few percent.  If we drop
terms proportional to $s_{13}^2$ in \eref{sol4}, we obtain the approximate sum
rule, 
\begin{equation*} \cos \delta \approx \frac {t_{23}(s^2_{12}-s^{\nu 2}_{12})}
{\sin 2\theta_{12}s_{13}}.  \end{equation*}
This sum rule can be written to leading order in $\theta_{13}$ as, \be
\frac{s^2_{12}-s^{\nu 2}_{12}} {2s_{12}c_{12}} \approx
\frac{\theta_{13}}{t_{23}}\cos \delta.  \label{sum5} \ee
If we write $\theta_{12}=\theta^{\nu}_{12}+\varepsilon_{12}$, then to leading
order in $\varepsilon_{12}$,
\begin{equation*} \frac{s^2_{12}-s^{\nu 2}_{12}} {2s_{12}c_{12}} \approx
\varepsilon_{12}.  \end{equation*}
Hence \eref{sum5} becomes, to leading order in $\varepsilon_{12}$ and
$\theta_{13}$, 
\be \theta_{12}-\theta^{\nu}_{12} \approx \frac{\theta_{13}}{t_{23}}\cos
\delta.  \label{sum6} \ee
If we write $\theta_{23}=\pi/4+\varepsilon_{23}$, then to leading order in
$\varepsilon_{23}$, $\varepsilon_{12}$ and $\theta_{13}$ we find,
\begin{equation} \theta_{12}-\theta^{\nu}_{12} \approx \theta_{13}\cos \delta,
\label{sum7} \end{equation}
which is the usual well-known leading-order solar sum
rule~\cite{King:2005bj,Masina:2005hf,Antusch:2005kw}.  The corrections to this
linearised sum rule are of order $\theta^2_{13}$, $\varepsilon^2_{12}$ and
$\theta_{13} \varepsilon_{23}$. For example, the second order correction from
the reactor angle is $\theta^2_{13}\sim (0.15)^2\sim 0.02\sim 2\%$.
The formulae preceding {\protect\eref{sum7}} may be used for a more accurate
approximate description of the sum rule. For example, {\protect\eref{sum6}}
could be used to better account for the atmospheric mixing angle deviating
from maximal.

\section{\label{app:klein}Neutrino mixing patterns from $\mathbb{Z}_2\times\mathbb{Z}_2$}

The possibility that the full $\mathbb{Z}_2\times\mathbb{Z}_2$ symmetry of the
neutrino mass term is a subgroup of the flavour symmetry was considered from a
bottom-up perspective in \refref{Hernandez:2012sk}. It is a simple extension of
the authors' previous constraints (see \refref{Hernandez:2012ra} for details):
each neutrino generator $S_i$ fixes one column of the mixing matrix using the
formulae,
\begin{equation} \begin{aligned}\cos^2\left(\frac{\pi d}{p}\right ) &=
\sin^2\left(\frac{\pi k}{m}\right )\left|U^\nu_{\alpha i}\right|^2, \\ 0 &=
\sin\left(\frac{2\pi k}{m}\right )(\left|U^\nu_{\beta i}\right|^2
- \left|U^\nu_{\gamma i}\right|^2), \end{aligned}\label{eq:app_HS_const}
  \end{equation}
where $k,m, d,p \in\mathbb{N}$ such that $0<k<m$ and $0<d<p$ with the
requirement that $k$ and $m$ ($d$ and $p$) are coprime, and
$\{\alpha,\beta,\gamma\}=\{e,\mu,\tau\}$.
Neglecting for the time being the case $m=2$ or $p=2$, the
constraint of unitarity implies that fixing two columns of the mixing matrix
using these formulae fully specifies the pattern. 
Each column is given by a permutation of $\{\sqrt{\eta},
\sqrt{\frac{1-\eta}{2}}, \sqrt{\frac{1-\eta}{2}}\}$, where the $\eta$ parameter
for the first and second column are given, respectively, by
\begin{align*} \eta_1 = \frac{\cos^2\left(\frac{\pi
d_1}{p_1}\right)}{\sin^2\left(\frac{\pi
k_1}{m_1}\right)}\qquad\text{and}\qquad\eta_2 = \frac{\cos^2\left(\frac{\pi
d_2}{p_2}\right)}{\sin^2\left(\frac{\pi k_2}{m_2}\right)}.  \end{align*}
We can generate the consistent mixing patterns by choosing those
$k_i, m_i, p_i$ and $d_i$ parameters which generate finite groups.
Considering only patterns related to von Dyck groups with irreducible triplets,
\ie A$_4$, S$_4$ and A$_5$, the result is the set of numbers listed in
\eref{eq:choices}. As we are assuming a single $\mathbb{Z}_m$ residual symmetry
in the charged-lepton sector, we must also take a common value of
$m$, \ie~$m_1=m_2$.

Up to row and column permutations, there are two ways of relatively aligning
the elements of the two fixed columns. In one case, we choose the unique
element fixed by the two real constraints to be in the same row, 
\begin{equation}  |U^\nu_{\alpha i}|^2 =  \left( \begin{matrix} \eta_1 & \eta_2 &
1-\eta_1-\eta_2 \\ \frac{1-\eta_1}{2} & \frac{1-\eta_2}{2} &
\frac{\eta_1+\eta_2}{2}\\ \frac{1-\eta_1}{2} & \frac{1-\eta_2}{2} &
\frac{\eta_1+\eta_2}{2} \end{matrix}\right). \label{PMNS_pattern}\end{equation}
In the other case, we choose these elements to be misaligned,
\begin{equation}  |U^\nu_{\alpha i}|^2 =  \left( \begin{matrix} \eta_1 &
\frac{1-\eta_2}{2} & \frac{1}{2}-\eta_1+\frac{\eta_2}{2} \\ \frac{1-\eta_1}{2}
& \eta_2 & \frac{1}{2}-\eta_2+\frac{\eta_1}{2}\\  \frac{1-\eta_1}{2} &
\frac{1-\eta_2}{2} & \frac{\eta_1+\eta_2}{2}
\end{matrix}\right).\label{PMNS_bad_pattern} \end{equation}
However, patterns of the form given in \eref{PMNS_bad_pattern} are not
possible if the residual symmetry in the charged-lepton sector has only a single
generator: it is the choice of $T_\alpha$ which specifies the row $\alpha$ of
the $\eta$ parameter. Under our assumption of a single cyclic charged-lepton
symmetry, $\mathbb{Z}_m$, we only need to consider patterns of the form
of \eref{PMNS_pattern} and its row and column permutations.

Up to now, we have refrained from discussing the case related to dihedral
groups where $\{m,p \}=\{2,N\}$. In fact, such a scenario does not give rise to
any new eligible neutrino patterns.  If $m$ or $p$ take the value $2$, the
symmetry constraints from the generators of this subgroup in
{\protect\eref{eq:app_HS_const}} do not necessarily fix a column of matrix
elements, leaving parts of the leading-order mixing matrix unspecified. 
Therefore, we will only consider the choice which does indeed fix a column
completely; it is given by $(m,p) = (N,2)$ with $N>2$ and yields
\[ U^\nu_{\alpha i} = 0\qquad \text{and}\qquad \left| U^\nu_{\beta i}\right | =
\left |U^\nu_{\gamma i}\right |.  \]
Then, the only way to fully specify the mixing matrix is to either apply this
constraint twice, or to apply this constraint in conjunction with one related
to A$_4$, S$_4$ or A$_5$.
By scanning through the row and column permutations of \eref{PMNS_pattern}, we
find that the only viable choices are (unsurprisingly) when the dihedral
constraint is applied once and is used to fix $U^\nu_{e3} = 0$; however, all
patterns of this type reproduce patterns already discussed previously.

%%%%%%%%%%%%%%%%%%%%%%%%%%%%%%


\begin{thebibliography}{99}

   %%%%Daya Bay
%\cite{An:2012eh}
\bibitem{An:2012eh}
  F.~P.~An {\it et al.}  [DAYA-BAY Collaboration],
 % ``Observation of electron-antineutrino disappearance at Daya Bay,''
  Phys.\ Rev.\ Lett.\  {\bf 108} (2012) 171803
  [arXiv:1203.1669];
  %%CITATION = ARXIV:1203.1669;%%
 %%%RENO
%\cite{Ahn:2012nd}
%\bibitem{Ahn:2012nd}
  J.~K.~Ahn {\it et al.}  [RENO Collaboration],
  %``Observation of Reactor Electron Antineutrino Disappearance in the RENO Experiment,''
  Phys.\ Rev.\ Lett.\  {\bf 108} (2012) 191802
  [arXiv:1204.0626].
  %%CITATION = ARXIV:1204.0626;%%
  
  %\cite{Gonzalez-Garcia:2014bfa}
\bibitem{Gonzalez-Garcia:2014bfa}
  M.~C.~Gonzalez-Garcia, M.~Maltoni and T.~Schwetz,
  %``Updated fit to three neutrino mixing: status of leptonic CP violation,''
  arXiv:1409.5439.
  %%CITATION = ARXIV:1409.5439;%%
  
  %\cite{Forero:2014bxa}
\bibitem{Forero:2014bxa}
  D.~V.~Forero, M.~Tortola and J.~W.~F.~Valle,
  %``Neutrino oscillations refitted,''
  arXiv:1405.7540.
  %%CITATION = ARXIV:1405.7540;%%
  %41 citations counted in INSPIRE as of 27 Oct 2014
  
  %\cite{Capozzi:2013csa}
\bibitem{Capozzi:2013csa}
  F.~Capozzi, G.~L.~Fogli, E.~Lisi, A.~Marrone, D.~Montanino and A.~Palazzo,
  %``Status of three-neutrino oscillation parameters, circa 2013,''
  Phys.\ Rev.\ D {\bf 89} (2014) 093018
  [arXiv:1312.2878].
  %%CITATION = ARXIV:1312.2878;%%
  %104 citations counted in INSPIRE as of 27 Oct 2014

%\cite{Harrison:2002er}
\bibitem{Harrison:2002er}
  P.~F.~Harrison, D.~H.~Perkins and W.~G.~Scott,
  %``Tri-bimaximal mixing and the neutrino oscillation data,''
  Phys.\ Lett.\ B {\bf 530} (2002) 167
  [hep-ph/0202074].
  %%CITATION = HEP-PH/0202074;%%
  
  %\cite{King:2009ap}
\bibitem{King:2009ap}
  S.~F.~King and C.~Luhn,
  %``On the origin of neutrino flavour symmetry,''
  JHEP {\bf 0910} (2009) 093
  [arXiv:0908.1897].
  %%CITATION = ARXIV:0908.1897;%%
  %54 citations counted in INSPIRE as of 05 Aug 2014

%\cite{King:2005bj}
\bibitem{King:2005bj}
  S.~F.~King,
  %``Predicting neutrino parameters from SO(3) family symmetry and quark-lepton unification,''
  JHEP {\bf 0508} (2005) 105
  [hep-ph/0506297].
  %%CITATION = HEP-PH/0506297;%%
  %203 citations counted in INSPIRE as of 05 Aug 2014

%\cite{King:2013eh}
\bibitem{King:2013eh}
  S.~F.~King and C.~Luhn,
  %``Neutrino Mass and Mixing with Discrete Symmetry,''
  Rept.\ Prog.\ Phys.\  {\bf 76} (2013) 056201
  [arXiv:1301.1340].
  %%CITATION = ARXIV:1301.1340;%%
  %123 citations counted in INSPIRE as of 05 Aug 2014
    
  %\cite{King:2014nza}
\bibitem{King:2014nza}
  S.~F.~King, A.~Merle, S.~Morisi, Y.~Shimizu and M.~Tanimoto,
  %``Neutrino Mass and Mixing: from Theory to Experiment,''
  New J.\ Phys.\  {\bf 16} (2014) 045018
  [arXiv:1402.4271].
  %%CITATION = ARXIV:1402.4271;%%
  %22 citations counted in INSPIRE as of 05 Aug 2014
  
  %\cite{King:2007pr}
\bibitem{King:2007pr}
  S.~F.~King,
  %``Parametrizing the lepton mixing matrix in terms of deviations from tri-bimaximal mixing,''
  Phys.\ Lett.\ B {\bf 659} (2008) 244
  [arXiv:0710.0530].
  %%CITATION = ARXIV:0710.0530;%%
  %93 citations counted in INSPIRE as of 05 Aug 2014

%\cite{Hernandez:2012ra}
\bibitem{Hernandez:2012ra}
  D.~Hernandez and A.~Y.~Smirnov,
  %``Lepton mixing and discrete symmetries,''
  Phys.\ Rev.\ D {\bf 86} (2012) 053014
  [arXiv:1204.0445].
  %%CITATION = ARXIV:1204.0445;%%
  %61 citations counted in INSPIRE as of 07 May 2014

%\cite{Hernandez:2012sk}
\bibitem{Hernandez:2012sk}
  D.~Hernandez and A.~Y.~Smirnov,
  %``Discrete symmetries and model-independent patterns of lepton mixing,''
  Phys.\ Rev.\ D {\bf 87} (2013) 5,  053005
  [arXiv:1212.2149].
  %%CITATION = ARXIV:1212.2149;%%
  %35 citations counted in INSPIRE as of 07 May 2014
 
  %\cite{Ballett:2013wya}
\bibitem{Ballett:2013wya}
  P.~Ballett, S.~F.~King, C.~Luhn, S.~Pascoli and M.~A.~Schmidt,
  %``Testing atmospheric mixing sum rules at precision neutrino facilities,''
  Phys.\ Rev.\ D {\bf 89} (2014) 016016
  [arXiv:1308.4314].
  %%CITATION = ARXIV:1308.4314;%%
  %6 citations counted in INSPIRE as of 07 May 2014


  %\cite{Shimizu:2011xg}
\bibitem{Shimizu:2011xg}
  Y.~Shimizu, M.~Tanimoto and A.~Watanabe,
  %``Breaking Tri-bimaximal Mixing and Large $\theta_{13}$,''
  Prog.\ Theor.\ Phys.\  {\bf 126} (2011) 81
  [arXiv:1105.2929].
  %%CITATION = ARXIV:1105.2929;%%
  %50 citations counted in INSPIRE as of 05 Aug 2014
  
  %\cite{King:2011zj}
\bibitem{King:2011zj}
  S.~F.~King and C.~Luhn,
  %``Trimaximal neutrino mixing from vacuum alignment in A4 and S4 models,''
  JHEP {\bf 1109} (2011) 042
  [arXiv:1107.5332].
  %%CITATION = ARXIV:1107.5332;%%
  %69 citations counted in INSPIRE as of 05 Aug 2014
 

  %\cite{Antusch:2011ic}
\bibitem{Antusch:2011ic}
  S.~Antusch, S.~F.~King, C.~Luhn and M.~Spinrath,
  %``Trimaximal mixing with predicted $\theta_{13}$ from a new type of constrained sequential dominance,''
  Nucl.\ Phys.\ B {\bf 856} (2012) 328
  [arXiv:1108.4278].
  %%CITATION = ARXIV:1108.4278;%%
  %64 citations counted in INSPIRE as of 05 Aug 2014
  
  %\cite{King:2013iva}
\bibitem{King:2013iva}
  S.~F.~King,
  %``Minimal predictive see-saw model with normal neutrino mass hierarchy,''
  JHEP {\bf 1307} (2013) 137
  [arXiv:1304.6264];
  %%CITATION = ARXIV:1304.6264;%%
  %10 citations counted in INSPIRE as of 05 Aug 2014
  %\cite{King:2013xba}
%\bibitem{King:2013xba}
  S.~F.~King,
  %``Minimal see-saw model predicting best fit lepton mixing angles,''
  Phys.\ Lett.\ B {\bf 724} (2013) 92
  [arXiv:1305.4846].
  %%CITATION = ARXIV:1305.4846;%%
  %8 citations counted in INSPIRE as of 05 Aug 2014
  
%\cite{Masina:2005hf}
\bibitem{Masina:2005hf}
  I.~Masina,
  %``A Maximal atmospheric mixing from a maximal CP violating phase,''
  Phys.\ Lett.\ B {\bf 633} (2006) 134
  [hep-ph/0508031].
  %%CITATION = HEP-PH/0508031;%%
  %70 citations counted in INSPIRE as of 05 Aug 2014

  %\cite{Antusch:2005kw}
\bibitem{Antusch:2005kw}
  S.~Antusch and S.~F.~King,
  %``Charged lepton corrections to neutrino mixing angles and CP phases revisited,''
  Phys.\ Lett.\ B {\bf 631} (2005) 42
  [hep-ph/0508044].
  %%CITATION = HEP-PH/0508044;%%
  %127 citations counted in INSPIRE as of 05 Aug 2014
  
%\cite{Antusch:2007rk}
\bibitem{Antusch:2007rk}
  S.~Antusch, P.~Huber, S.~F.~King and T.~Schwetz,
  %``Neutrino mixing sum rules and oscillation experiments,''
  JHEP {\bf 0704} (2007) 060
  [hep-ph/0702286].
  %%CITATION = HEP-PH/0702286;%%  
  %61 citations counted in INSPIRE as of 05 Aug 2014
 
  %\cite{Marzocca:2013cr}
\bibitem{Marzocca:2013cr}
  D.~Marzocca, S.~T.~Petcov, A.~Romanino and M.~C.~Sevilla,
  %``Nonzero |U_e3| from Charged Lepton Corrections and the Atmospheric Neutrino Mixing Angle,''
  JHEP {\bf 1305} (2013) 073
  [arXiv:1302.0423].
  %%CITATION = ARXIV:1302.0423;%%
  %15 citations counted in INSPIRE as of 05 Aug 2014

  %\cite{Petcov:2014laa}
\bibitem{Petcov:2014laa}
  S.~T.~Petcov,
  %``Predicting the Values of the Leptonic CP Violation Phases,''
  arXiv:1405.6006;
  %%CITATION = ARXIV:1405.6006;%%
  %1 citations counted in INSPIRE as of 05 Aug 2014

 %\cite{Barger:1998ta}
\bibitem{Barger:1998ta}
M.~Fukugita, M.~Tanimoto and T.~Yanagida,
  %``Atmospheric neutrino oscillation and a phenomenological lepton mass matrix,''
  Phys.\ Rev.\ D {\bf 57} (1998) 4429
  [hep-ph/9709388];
%
  V.~D.~Barger, S.~Pakvasa, T.~J.~Weiler and K.~Whisnant,
  %``Bimaximal mixing of three neutrinos,''
  Phys.\ Lett.\ B {\bf 437} (1998) 107
  [hep-ph/9806387];
  %%CITATION = HEP-PH/9806387;%%
  %444 citations counted in INSPIRE as of 03 Jul 2014
S.~Davidson and S.~F.~King,
  %``Bimaximal neutrino mixing in the MSSM with a single right-handed neutrino,''
  Phys.\ Lett.\ B {\bf 445} (1998) 191
  [hep-ph/9808296].

\bibitem{Datta:2003qg}
  A.~Datta, F.~S.~Ling and P.~Ramond,
  %``Correlated hierarchy, Dirac masses and large mixing angles,''
  Nucl.\ Phys.\ B {\bf 671} (2003) 383
  [hep-ph/0306002];
 %
  L.~L.~Everett and A.~J.~Stuart,
  %``Icosahedral (A(5)) Family Symmetry and the Golden Ratio Prediction for Solar Neutrino Mixing,''
  Phys.\ Rev.\ D {\bf 79} (2009) 085005
  [arXiv:0812.1057];
%
  F.~Feruglio and A.~Paris,
  %``The Golden Ratio Prediction for the Solar Angle from a Natural Model with A5 Flavour Symmetry,''
  JHEP {\bf 1103} (2011) 101
  [arXiv:1101.0393].

%\cite{Lam:2011ag}
\bibitem{Lam:2011ag}
  C.~S.~Lam,
  %``Group Theory and Dynamics of Neutrino Mixing,''
  Phys.\ Rev.\ D {\bf 83} (2011) 113002
  [arXiv:1104.0055].
  %%CITATION = ARXIV:1104.0055;%%
  %28 citations counted in INSPIRE as of 20 Aug 2014

%\cite{deAdelhartToorop:2011re}
\bibitem{deAdelhartToorop:2011re}
  R.~de Adelhart Toorop, F.~Feruglio and C.~Hagedorn,
  %``Finite Modular Groups and Lepton Mixing,''
  Nucl.\ Phys.\ B {\bf 858} (2012) 437
  [arXiv:1112.1340].
  %%CITATION = ARXIV:1112.1340;%%
  %67 citations counted in INSPIRE as of 20 Aug 2014

\bibitem{Rodejohann:2008ir}
  W.~Rodejohann,
  %``Unified Parametrization for Quark and Lepton Mixing Angles,''
  Phys.\ Lett.\ B {\bf 671} (2009) 267
  [arXiv:0810.5239];
 %
  A.~Adulpravitchai, A.~Blum and W.~Rodejohann,
  %``Golden Ratio Prediction for Solar Neutrino Mixing,''
  New J.\ Phys.\  {\bf 11} (2009) 063026
  [arXiv:0903.0531].

%\cite{Albright:2010ap}
\bibitem{Albright:2010ap}
  C.~H.~Albright, A.~Dueck and W.~Rodejohann,
  %``Possible Alternatives to Tri-bimaximal Mixing,''
  Eur.\ Phys.\ J.\ C {\bf 70} (2010) 1099
  [arXiv:1004.2798].
  %%CITATION = ARXIV:1004.2798;%%
  %78 citations counted in INSPIRE as of 05 Aug 2014
  
%\cite{Petcov:2001sy}
\bibitem{Petcov:2001sy}
  S.~T.~Petcov and M.~Piai,
  %``The LMA MSW solution of the solar neutrino problem, inverted neutrino mass hierarchy and reactor neutrino experiments,''
  Phys.\ Lett.\ B {\bf 533} (2002) 94
  [hep-ph/0112074].
  %%CITATION = HEP-PH/0112074;%%
  %98 citations counted in INSPIRE as of 07 May 2014

%\cite{Qian:2012xh}
\bibitem{Qian:2012xh}
  X.~Qian, D.~A.~Dwyer, R.~D.~McKeown, P.~Vogel, W.~Wang and C.~Zhang,
  %``Mass Hierarchy Resolution in Reactor Anti-neutrino Experiments: Parameter Degeneracies and Detector Energy Response,''
  Phys.\ Rev.\ D {\bf 87} (2013) 3,  033005
  [arXiv:1208.1551].
  %%CITATION = ARXIV:1208.1551;%%
  %34 citations counted in INSPIRE as of 07 May 2014

%\cite{Ge:2012wj}
\bibitem{Ge:2012wj}
  S.~-F.~Ge, K.~Hagiwara, N.~Okamura and Y.~Takaesu,
  %``Determination of mass hierarchy with medium baseline reactor neutrino experiments,''
  JHEP {\bf 1305} (2013) 131
  [arXiv:1210.8141].
  %%CITATION = ARXIV:1210.8141;%%
  %21 citations counted in INSPIRE as of 07 May 2014

%\cite{Li:2013zyd}
\bibitem{Li:2013zyd}
  Y.~-F.~Li, J.~Cao, Y.~Wang and L.~Zhan,
  %``Unambiguous Determination of the Neutrino Mass Hierarchy Using Reactor Neutrinos,''
  Phys.\ Rev.\ D {\bf 88} (2013) 1,  013008
  [arXiv:1303.6733].
  %%CITATION = ARXIV:1303.6733;%%
  %44 citations counted in INSPIRE as of 07 May 2014

\bibitem{Wang:2013aa}
  W. Wang,
  Talk presented at Invisibles '13 at Lumley Castle, Durham on 16/07/2013. 

%\cite{Park:2014jda}
\bibitem{Park:2014jda}
  J.~Park,
  %``Recent results from RENO,''
  AIP Conf.\ Proc.\  {\bf 1604} (2014) 421.
  %%CITATION = APCPC,1604,421;%%
%\cite{Park:2014sja}

\bibitem{Park:2014sja}
  J.~Park,
  %``Study of Neutrino Mass Hierarchy with RENO-50,''
  PoS Neutel {\bf 2013} (2013) 076.
  %%CITATION = POSCI,Neutel2013,076;%%

%\cite{Adams:2013qkq}
\bibitem{Adams:2013qkq}
  C.~Adams {\it et al.}  [LBNE Collaboration],
  %``The Long-Baseline Neutrino Experiment: Exploring Fundamental Symmetries of the Universe,''
  arXiv:1307.7335.
  %%CITATION = ARXIV:1307.7335;%%
  %88 citations counted in INSPIRE as of 23 Oct 2014

%\cite{::2013kaa}
\bibitem{Agarwalla:2013kaa}
  S.~Agarwalla : {\it et al.}  [LAGUNA-LBNO Collaboration],
  %``The mass-hierarchy and CP violation discovery reach of the LBNO long-baseline neutrino experiment,''
  arXiv:1312.6520.
  %%CITATION = ARXIV:1312.6520;%%
  %4 citations counted in INSPIRE as of 12 May 2014

%\cite{Ballett:2014uia}
\bibitem{Ballett:2014uia}
  P.~Ballett, S.~F.~King, C.~Luhn, S.~Pascoli and M.~A.~Schmidt,
  %``Precision measurements of {\theta}12 for testing models of discrete leptonic flavour symmetries,''
  arXiv:1406.0308.
  %%CITATION = ARXIV:1406.0308;%%
  %1 citations counted in INSPIRE as of 27 Oct 2014

 %\cite{Gollu:2013yla}
\bibitem{Gollu:2013yla}
  S.~Gollu, K.~N.~Deepthi and R.~Mohanta,
  %``Charged lepton correction to tri-bimaximal lepton mixing and its implications to neutrino phenomenology,''
  Mod.\ Phys.\ Lett.\ A {\bf 28} (2013) 31,  1350131
  [arXiv:1303.3393].
  %%CITATION = ARXIV:1303.3393;%%
  %6 citations counted in INSPIRE as of 02 Sep 2014

%\cite{M.:2014kca}
\bibitem{Sruthilaya:2014kca}
  Sruthilaya~M., Soumya~C., K.~N.~Deepthi and R.~Mohanta,
  %``Predicting Leptonic CP phase by considering deviations in charged lepton and neutrino sectors,''
  arXiv:1408.4392.
  %%CITATION = ARXIV:1408.4392;%%

%\cite{Girardi:2014faa}
\bibitem{Girardi:2014faa}
  I.~Girardi, S.~T.~Petcov and A.~V.~Titov,
  %``Determining the Dirac CP Violation Phase in the Neutrino Mixing Matrix from Sum Rules,''
  arXiv:1410.8056.
  %%CITATION = ARXIV:1410.8056;%%
   
%\cite{Agashe:2014kda}
\bibitem{Agashe:2014kda}
  K.~A.~Olive {\it et al.}  [Particle Data Group Collaboration],
  %``Review of Particle Physics,''
  Chin.\ Phys.\ C {\bf 38} (2014) 090001.
  %%CITATION = CHPHD,C38,090001;%%
  %27 citations counted in INSPIRE as of 29 Sep 2014

%\cite{Coloma:2012wq}
\bibitem{Coloma:2012wq}
  P.~Coloma, A.~Donini, E.~Fernandez-Martinez and P.~Hernandez,
  %``Precision on leptonic mixing parameters at future neutrino oscillation experiments,''
  JHEP {\bf 1206} (2012) 073
  [arXiv:1203.5651].
  %%CITATION = ARXIV:1203.5651;%%
  %27 citations counted in INSPIRE as of 06 Nov 2014

%\cite{deGouvea:2013onf}
\bibitem{deGouvea:2013onf}
  A.~de Gouvea {\it et al.}  [Intensity Frontier Neutrino Working Group Collaboration],
  %``Working Group Report: Neutrinos,''
  arXiv:1310.4340.
  %%CITATION = ARXIV:1310.4340;%%
  %39 citations counted in INSPIRE as of 07 Nov 2014

%\cite{Fonseca:2014koa}
\bibitem{Fonseca:2014koa}
  R.~M.~Fonseca and W.~Grimus,
  %``Classification of lepton mixing matrices from finite residual symmetries,''
  JHEP {\bf 1409} (2014) 033
  [arXiv:1405.3678].
  %%CITATION = ARXIV:1405.3678;%%
  %7 citations counted in INSPIRE as of 29 Sep 2014

%\cite{Blum:2007jz}
\bibitem{Blum:2007jz}
  A.~Blum, C.~Hagedorn and M.~Lindner,
  %``Fermion Masses and Mixings from Dihedral Flavor Symmetries with Preserved Subgroups,''
  Phys.\ Rev.\ D {\bf 77} (2008) 076004
  [arXiv:0709.3450].
  %%CITATION = ARXIV:0709.3450;%%
  %90 citations counted in INSPIRE as of 01 Oct 2014

%\cite{Capozzi:2013psa}
\bibitem{Capozzi:2013psa}
  F.~Capozzi, E.~Lisi and A.~Marrone,
  %``Neutrino mass hierarchy and electron neutrino oscillation parameters with one hundred thousand reactor events,''
  Phys.\ Rev.\ D {\bf 89} (2014) 013001
  [arXiv:1309.1638].
  %%CITATION = ARXIV:1309.1638;%%
  %12 citations counted in INSPIRE as of 12 May 2014

%\cite{Grassi:2014hxa}
\bibitem{Grassi:2014hxa}
  M.~Grassi, J.~Evslin, E.~Ciuffoli and X.~Zhang,
  %``Showering Cosmogenic Muons in A Large Liquid Scintillator,''
  arXiv:1401.7796.
  %%CITATION = ARXIV:1401.7796;%%
  %1 citations counted in INSPIRE as of 16 Sep 2014

%\cite{Stahl:2012exa}
\bibitem{Stahl:2012exa}
  A.~Stahl, C.~Wiebusch, A.~M.~Guler, M.~Kamiscioglu, R.~Sever, A.~U.~Yilmazer, C.~Gunes and D.~Yilmaz {\it et al.},
  %``Expression of Interest for a very long baseline neutrino oscillation experiment (LBNO),''
  CERN-SPSC-2012-021.
  %%CITATION = CERN-SPSC-2012-021;%%
  %32 citations counted in INSPIRE as of 12 May 2014

%\cite{Longhin:2012ae}
\bibitem{Longhin:2012ae}
  A.~Longhin,
  %``Optimization of neutrino beams for underground sites in Europe,''
  arXiv:1206.4294.
  %%CITATION = ARXIV:1206.4294;%%
  %3 citations counted in INSPIRE as of 12 May 2014

%\cite{Huber:2004ka}
\bibitem{Huber:2004ka}
  P.~Huber, M.~Lindner and W.~Winter,
  %``Simulation of long-baseline neutrino oscillation experiments with GLoBES (General Long Baseline Experiment Simulator),''
  Comput.\ Phys.\ Commun.\  {\bf 167} (2005) 195
  [hep-ph/0407333].
  %%CITATION = HEP-PH/0407333;%%
  %250 citations counted in INSPIRE as of 12 May 2014

%\cite{Huber:2007ji}
\bibitem{Huber:2007ji}
  P.~Huber, J.~Kopp, M.~Lindner, M.~Rolinec and W.~Winter,
  %``New features in the simulation of neutrino oscillation experiments with GLoBES 3.0: General Long Baseline Experiment Simulator,''
  Comput.\ Phys.\ Commun.\  {\bf 177} (2007) 432
  [hep-ph/0701187].
  %%CITATION = HEP-PH/0701187;%%
  %193 citations counted in INSPIRE as of 12 May 2014

   %\cite{King:2012in}
\bibitem{King:2012in}
  S.~F.~King, C.~Luhn and A.~J.~Stuart,
  %``A Grand Delta(96) x SU(5) Flavour Model,''
  Nucl.\ Phys.\ B {\bf 867} (2013) 203
  [arXiv:1207.5741].
  %%CITATION = ARXIV:1207.5741;%%
  %29 citations counted in INSPIRE as of 22 Apr 2014

%\cite{Schmidt:2006rb}
\bibitem{Schmidt:2006rb}
  M.~A.~Schmidt and A.~Y.~Smirnov,
  %``Quark Lepton Complementarity and Renormalization Group Effects,''
  Phys.\ Rev.\ D {\bf 74} (2006) 113003
  [hep-ph/0607232].
  %%CITATION = HEP-PH/0607232;%%
  %69 citations counted in INSPIRE as of 15 Oct 2014

%\cite{Boudjemaa:2008jf}
\bibitem{Boudjemaa:2008jf}
  S.~Boudjemaa and S.~F.~King,
  %``Deviations from Tri-bimaximal Mixing: Charged Lepton Corrections and Renormalization Group Running,''
  Phys.\ Rev.\ D {\bf 79} (2009) 033001
  [arXiv:0808.2782].
  %%CITATION = ARXIV:0808.2782;%%
  %59 citations counted in INSPIRE as of 15 Oct 2014

%\cite{Antusch:2003kp}
\bibitem{Antusch:2003kp}
  S.~Antusch, J.~Kersten, M.~Lindner and M.~Ratz,
  %``Running neutrino masses, mixings and CP phases: Analytical results and phenomenological consequences,''
  Nucl.\ Phys.\ B {\bf 674} (2003) 401
  [hep-ph/0305273].
  %%CITATION = HEP-PH/0305273;%%
  %247 citations counted in INSPIRE as of 29 Apr 2014


\end{thebibliography}
\end{document}